\pdfoutput=1

\documentclass[lettersize,journal]{IEEEtran}
\usepackage{amsmath,amsfonts}
\usepackage{algorithmic}
\usepackage{algorithm}
\usepackage{array}
\usepackage[caption=false,font=normalsize,labelfont=sf,textfont=sf]{subfig}
\usepackage{textcomp}
\usepackage{stfloats}
\usepackage{url}
\usepackage{verbatim}
\usepackage{graphicx}
\usepackage{cite}
\usepackage{multirow}
\graphicspath{{Figures/}}
\usepackage{url}
\usepackage{xcolor}

\begin{document}

\title{Is it Possible to Generate Irreversible PolyProtected Templates\\ from Face Embeddings using System-Specific Keys?}

\author{Vedrana~Krivoku\'ca~Hahn, J\'er\'emy Maceiras, S\'ebastien~Marcel,~\IEEEmembership{Senior Member,~IEEE}% <-this % stops a space
\thanks{All authors are with Idiap Research Institute in Martigny, Switzerland.}}

\maketitle

\begin{abstract}

This work aims to answer the question of whether it is possible to generate \textit{irreversible} protected templates when the PolyProtect biometric template protection method is applied to face embeddings using \textit{system}-specific keys (i.e., the same $C$ and $E$ parameters, which define the transform, are applied to all subjects' face embeddings), instead of the traditional \textit{subject}-specific keys (i.e., each subject has their own $C$ and $E$ parameters).  This is important for determining whether we can perform de-duplication of face identities in the PolyProtected domain, which is not possible in the subject-specific key scenario due to the clash with PolyProtect's unlinkability property (i.e., one could generate multiple protected templates belonging to the same identity, using different $C$ and $E$ parameters, such that those templates cannot be linked to each other).  We present experiments\footnote{Reproducible using our open-source code: \textit{Link upon paper acceptance}.} to prove that there exist at least three ways of systematically selecting system-specific keys that produce irreversible PolyProtected templates: (i) from pre-selected subject-specific keys, (ii) by applying a previously proposed key selection algorithm to random vectors, and (iii) by approximating a ``good'' $C$/$E$ pair distribution from which system-specific keys can be constructed.  Our findings thus point to the conclusion that it is, indeed, possible to safely operate PolyProtect in the system-specific key scenario without degrading the template protection potential.  This opens up the possibility for identity de-duplication in the PolyProtected domain. 
    
\end{abstract}


\section{Introduction}
\label{sec:introduction}

\IEEEPARstart{F}aces are unique, face images are easy to capture, and we have access to highly accurate (and often free) face recognition algorithms as well as a variety of face image datasets.  It is no wonder, therefore, that automated face recognition technologies have become ubiquitous tools for human identity checks in many applications, ranging from personal device or account management (e.g., accessing a laptop or smartphone, logging in to online services such as banking) to large-scale identity checks (e.g., passport verification at electronic airport gates, surveillance in crowds).  While the convenience and improved security provided by this form of authentication (compared to traditional methods like passwords or manual identity checks) are evident, the price we pay is our privacy.  The more our face data is harvested, the greater the chances of this data being leaked and used for nefarious purposes, such as impersonation through deepfake or spoofing attacks, or tracking across different applications in which the same faces are enrolled.  So, to ensure that we can benefit from face recognition technologies without jeopardising our privacy, it is important to protect the employed face data.  

The protection of face (and other biometric) data falls within the domain of Biometric Template Protection (BTP).  The aim of BTP is to create a \textit{protected template} from a biometric feature vector (unprotected template), such that it is impossible to ``invert'' the protected template to recover the unprotected one.  In this paper, we consider BTP in the context of face recognition systems.  Modern face recognition systems avail of deep-learning-based ``models'' that have been trained to convert face images to fixed-length numerical representations (feature vectors) called ``embeddings''.  Face embeddings have been shown to be invertible \cite{bg26}, meaning they can be used to recover an approximation of the underlying face image \cite{z16, c17, m19, s22}, and soft biometric attributes (e.g., sex, race, age, hair colour) can be extracted from these representations \cite{f20, t20}.  So, face embeddings must be protected (via BTP), to prevent recovery of the original face information and thus preserve the privacy of the people enrolled in the face recognition system.  Face BTP methods come in two main flavours:  Handcrafted (designed by humans) and Learned (learned via neural networks).

Handcrafted BTP methods include: feature transformations \cite{d26} (transforming a biometric template to a protected feature space, e.g., \cite{d19, kh22}); biometric cryptosystems \cite{r26} (usually ``binding'' a biometric template with an external key, e.g., \cite{g19, r22}); and homomorphic encryption \cite{b26} (matching reference and probe templates in the encrypted domain, e.g., \cite{m17, b18, e22}).  Learned BTP methods \cite{kh26} include: training a neural network to transform a biometric template to a pre-defined random code (e.g., \cite{p16, j18}); or training a neural network to learn its own representation of a protected template (e.g., \cite{t19, p21, m21}).  Learned BTP methods have the potential for more complex transforms that generate more secure protected templates, as we are not limited by the know-how of human engineers; however, handcrafted methods are easier to use in practice, since they can more readily be integrated into existing biometric systems (e.g., as a module after the feature extractor), and they tend to be easier to evaluate since they are designed by humans and thus have more explicit properties compared to black-box neural-network-learned BTP algorithms.  For these reasons, we chose to focus on \textit{handcrafted} BTP methods, and we selected PolyProtect \cite{kh22} as our case-study. 

PolyProtect transforms face embeddings into protected templates using multivariate polynomials, whose coefficients ($C$) and exponents ($E$) are external parameters.  A polynomial is applied to consecutive sets of embedding elements, where the amount of overlap between the sets is a tunable parameter.  We selected PolyProtect for the same reasons as those outlined in \cite{khm26}, i.e., it is easy and lightweight to implement, modular (can be integrated into existing face recognition systems), satisfies the main BTP criteria (recognition accuracy, irreversibility, unlinkability) \cite{kh22, s25, khm26}, and comes with open-source code.\footnote{\url{https://gitlab.idiap.ch/bob/bob.paper.polyprotect_2021}}  More specifically, our work builds upon the findings from \cite{khm26}, so it falls within the context of the same global aim, which is to develop a privacy-preserving, open-source face identification system for humanitarian aid distribution.  PolyProtect thus remains the best choice of BTP method as it most closely aligns with our requirements.  

The work in \cite{khm26} presented a deeper dive into the ``irreversibility'' of PolyProtect.  One of the main contributions was a new \textit{key selection algorithm}, which ensures that the $C$ and $E$ parameters (``keys'') selected for PolyProtect's transformation polynomials result in the generation of \textit{irreversible} protected templates.  The need for this algorithm was motivated by the finding that there exists a more powerful worst-case attacker than that assumed in the original PolyProtect work \cite{kh22}, whereby a cosine-distance-based numerical solver is used to invert a PolyProtected face template produced from \textit{randomly generated} $C$ and $E$ to recover a very close approximation (in terms of cosine distance) to the original face embedding.  It was thus recommended that $C$ and $E$ be chosen using the proposed key selection algorithm, which was shown to significantly improve the irreversibility of PolyProtected templates compared to when these parameters are generated in a purely random fashion.  This investigation was conducted in the scenario where $C$ and $E$ are \textit{subject-specific}, i.e., each subject (identity) enrolled in the face recognition system has their \textit{own} (unique) set of these parameters.  This makes sense, as it is in line with how PolyProtect was designed to operate.  

We now face the following problem.  The ultimate goal of our work is to investigate whether it is possible to perform \textit{de-duplication} of face templates in the \textit{protected} domain, in order to check whether the same person has enrolled more than once in the same face recognition system (for the purposes of ``double-dipping'', or accessing more than their fair share of resources in a humanitarian aid distribution setting), while ensuring that the face data remains protected during this (post-enrollment) process.  In this case, employing subject-specific keys for PolyProtect would not work, since this would conflict with the \textit{unlinkability} property of this BTP method, i.e., since it is possible to generate \textit{N} unlinkable (non-matching) protected templates from the same face identity using \textit{N} different ``keys'' (sets of $C$, $E$ parameters), then it is effectively possible to simulate \textit{N} different face identities in the protected domain, making it impossible to detect multiple enrollments of this particular face identity.  So, the aim of this paper is to investigate the use of \textit{system-specific} keys instead, whereby all subjects would be enrolled in the face recognition system using the \textit{same} $C$ and $E$ parameters to generate their PolyProtected templates.  In this case, all face embeddings would be transformed in the same way, making it possible to perform de-duplication in the protected domain.        

\begin{figure}[!h]
\centering
\includegraphics[width=0.8\columnwidth]{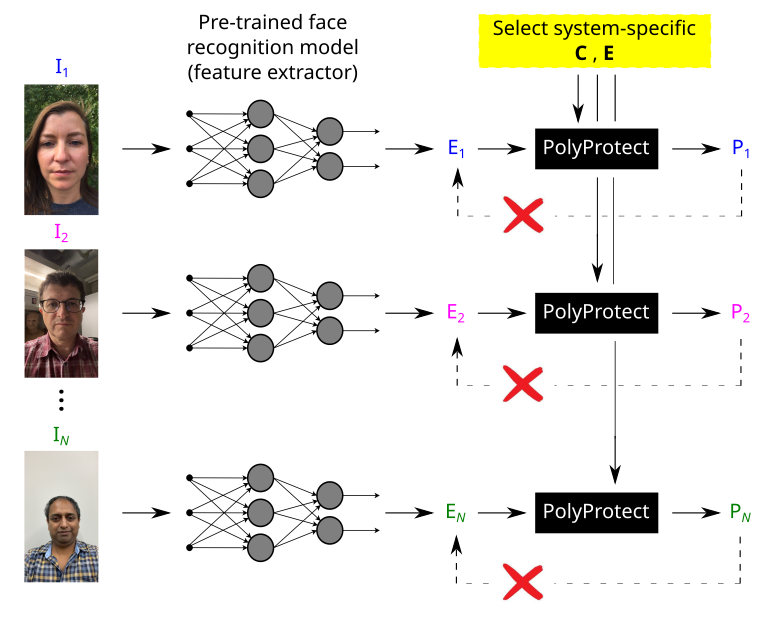}%
\caption{We investigate the selection of \textit{system}-specific keys ($C$, $E$) for the PolyProtect BTP method, such that the protected templates ($P_1$, ..., $P_N$), generated from the face embeddings ($E_1$, ..., $E_N$) of all identities enrolled in the face recognition system ($I_1$, ..., $I_N$) are \textit{irreversible}, i.e., it is impossible to recover $E_i$ from its corresponding $P_i$.  (Face images from SOTERIA \cite{r24}.)}
\label{fig:teaser}
\end{figure}

Our main concern with a \textit{system}-specific key approach relates to the effect that the selected keys may have on the \textit{irreversibility} of the protected templates.  In particular, the following issue arises: How do we select a single system-specific key (i.e., a set of $C$ and $E$ parameters) that would ``work'' for all subjects, meaning that the key would lead to the generation of \textit{irreversible} protected templates for everyone enrolled in the face recognition system?  This question was born out of the findings in \cite{khm26}, where it was recommended that PolyProtect's keys \textit{not} be selected randomly, since ``bad'' keys may result in \textit{reversible} (invertible) protected templates; however, the proposed key selection algorithm, which was shown to be capable of selecting ``good'' keys that lead to \textit{irreversible} protected templates, was investigated only in the \textit{subject}-specific key scenario, where the keys were chosen separately for each subject.  So, we asked ourselves the following question: Could the \textit{same key selection algorithm} be used to select a single \textit{system}-specific key that would ensure the \textit{irreversibility} of all subjects' protected templates?  In other words, is the proposed key selection algorithm \textit{generalizable} to the system-specific key scenario, or is it limited to the selection of subject-specific keys?  This investigation is important for understanding whether or not PolyProtect can operate in the system-specific key scenario, in order to enable de-duplication in the protected domain while ensuring the irreversibility of the protected templates and thus protecting the underlying face data.  This is the focus of our paper, illustrated in Fig. \ref{fig:teaser}.  

Our main contributions are the following:

\begin{enumerate}

\item We show that PolyProtect ``keys'' (sets of $C$ and $E$ parameters) chosen for specific subjects' face embeddings using the key selection algorithm from \cite{khm26} are \textit{generalizable} to different subjects.  In other words, a key chosen for one subject's face embedding can be used to generate an irreversible PolyProtected template from another subject's face embedding.  This suggests that the key selection algorithm is \textit{not limited} to the subject-specific scenario, meaning that it seems suitable for the selection of system-specific keys.
	
\item We demonstrate that applying the key selection algorithm from \cite{khm26} to \textit{random vectors}, whose values lie in the same range as the target face embeddings, then applying the selected keys ($C$, $E$) to the face embeddings, still results in the generation of irreversible PolyProtected templates.  This implies that the key selection algorithm does not strictly depend on the face embeddings themselves, but rather on the range in which the embedding values lie.  Consequently, we show that ``good'' keys can actually be chosen independently of the face embeddings to which they will be applied (to generate the PolyProtected templates).  This further confirms the generalizability of the key selection algorithm from \cite{khm26} towards the system-specific key scenario.

\item We show that it is possible to use the keys selected on random vectors (from 2) to establish a ``good'' key \textit{distribution}, consisting of all the $C$ values chosen for each possible $E$ value.  We then select ($C_i$, $E_i$) pairs from this distribution to make up 5-pair ($C$, $E$) sequences that represent the keys, instead of pre-generating the full sequences and picking the keys directly from this list (as in 1 or 2).  We demonstrate that this approach, too, allows us to choose keys that lead to the generation of irreversible PolyProtected templates.  So, we confirm the existence of some set of ``good'' keys, including individual $C$/$E$ pairs, which the key selection algorithm from \cite{khm26} is able to identify, and which work in both the subject- \textit{and} system-specific PolyProtect scenarios.  

\end{enumerate}

The remainder of this paper is structured as follows.  Section \ref{sec:background} presents the background required to understand the work in this paper, namely the PolyProtect BTP method \cite{kh22} that is our focus and the key selection algorithm \cite{khm26} that we investigate in the system-specific key scenario.  Section \ref{sec:experiments} then dives into the main contributions of this work, presenting three experiments that investigate the generalizability of the key selection algorithm to the system-specific key scenario: the use of subject-specific keys to generate irreversible PolyProtected templates for different subjects, the use of random vectors instead of face embeddings to select system-specific keys that produce irreversible PolyProtected templates for all subjects, and the selection of system-specific keys directly from a ``good'' $C$/$E$ pair distribution.  Section \ref{sec:conclusion} wraps up the findings from this paper and proposes directions for future work.

\section{Background}
\label{sec:background}

This section explains the background needed to understand the contributions presented in this paper.  Firstly, Section \ref{subsec:polyprotect} presents an overview of PolyProtect, the BTP method that we have chosen to focus on for the reasons outlined in Section \ref{sec:introduction}.  Then, Section \ref{subsec:ks_alg} briefly describes the key selection algorithm from \cite{khm26}, which was proposed as a means of choosing PolyProtect's keys ($C$ and $E$ parameters) such that the resulting protected template would be \textit{irreversible} under the worst-case inversion attack based on a cosine-distance numerical solver.  In this paper we evaluate the efficacy of this key selection algorithm in the new context of \textit{system}-specific keys (as opposed to the \textit{subject}-specific keys used in \cite{khm26}), which is an important step towards our ultimate goal of performing identity de-duplication in the protected domain.

\subsection{PolyProtect}
\label{subsec:polyprotect}

PolyProtect, proposed in \cite{kh22} as a BTP method for face embeddings, works as follows.  Let $V = [v_1, v_2, ..., v_n]$ denote an $n$-dimensional face embedding. PolyProtect transforms $V$ into a \textit{protected} template, $P = [p_1, p_2, ..., p_k]$ ($k < n$), by mapping sets of $m$ consecutive elements from $V$ to single elements in $P$ via multivariate polynomials defined by $m$ coefficients, $C = [c_1, c_2, ..., c_m]$, and exponents, $E = [e_1, e_2, ..., e_m]$.  

The first $m$ elements in $V$ (i.e., $v_1, v_2, ..., v_m$) are transformed into the first element in $P$ (i.e., $p_1$) via Eq. (1):

\begin{equation}
p_1 = c_{1}v_{1}^{e_1} + c_{2}v_{2}^{e_2} + ... + c_{m}v_{m}^{e_m}
\end{equation}

The elements of $V$ used to generate $p_2$ depend on the chosen amount of \textit{overlap} between successive sets of elements. The minimum overlap is 0, in which case the elements of $V$ in each set would be unique, and the maximum is $m - 1$, in which case successive sets would share $m - 1$ elements. Eqs. (2) and (3) define the mapping from $V$ to $p_2$ for overlaps of 0 and $m - 1$, respectively:

\begin{equation}
p_2 = c_{1}v_{m+1}^{e_1} + c_{2}v_{m+2}^{e_2} + ... + c_{m}v_{m+m}^{e_m}	
\end{equation}

\begin{equation}
p_2 = c_{1}v_{2}^{e_1} + c_{2}v_{3}^{e_2} + ... + c_{m}v_{m+1}^{e_m}	
\end{equation}

The remaining elements in $P$ (i.e., $p_3, ..., p_k$) are generated in a similar manner, until all the elements in $V$ have been used up. If the last set of elements is incomplete because the dimensionality of $V$ is not divisible by the required number of sets (defined by $m$ and the amount of overlap), $V$ is padded by a sufficient number of zeros to complete the last set.

\begin{figure}[!h]
\centering
\includegraphics[width=\columnwidth]{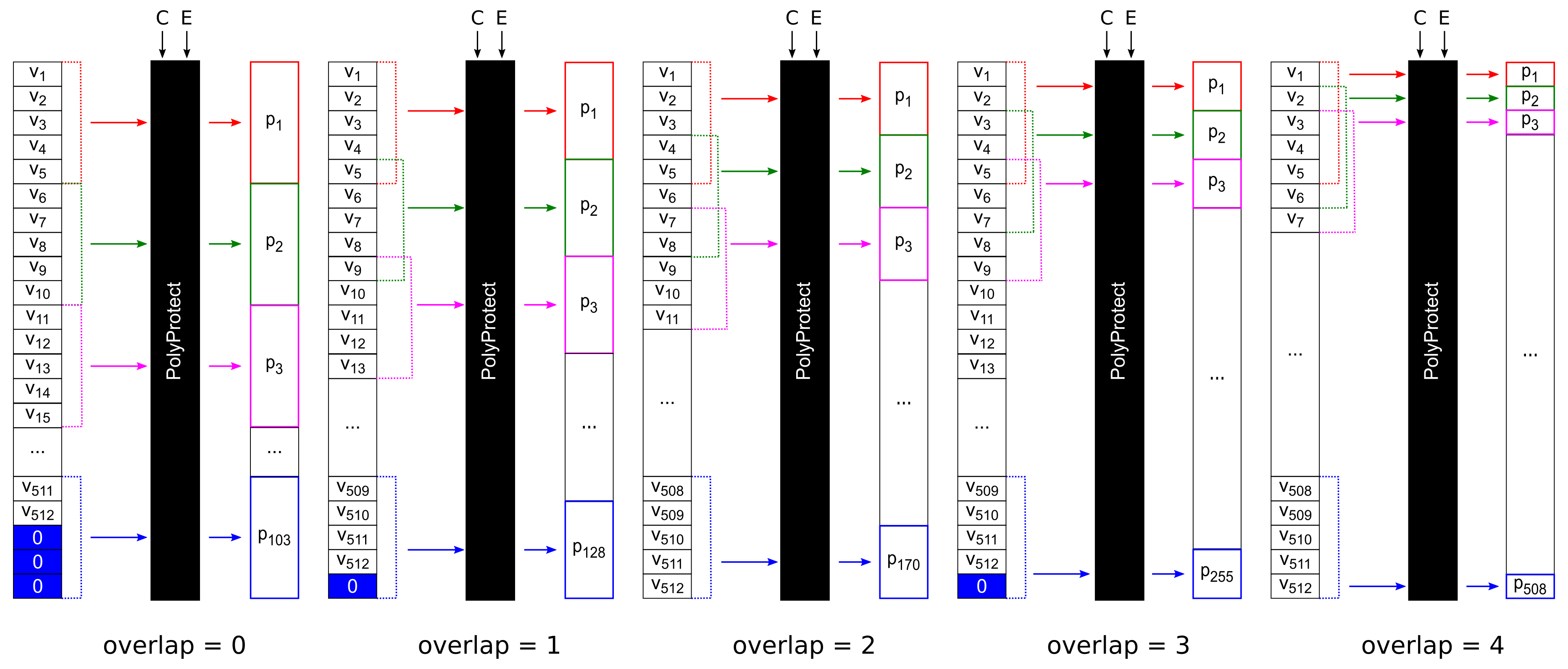}%
\caption{Mapping 512-dimensional $V$ to $P$ via PolyProtect, using $C = [c_1, c_2, ..., c_5]$ and $E = [e_1, e_2, ..., e_5]$, for different amounts of overlap.}
\label{fig:polyprotect}
\end{figure}

The analysis presented in this paper will be based on 512-dimensional face embeddings (see Section \ref{subsec:set_up}), so Fig. \ref{fig:polyprotect} illustrates the transformation from a 512-dimensional $V$ to $P$, for overlaps of 0 to 4, when $m = 5$.  It is evident that the dimensionality of $P$ is influenced by the amount of overlap used in the $V \rightarrow P$ mapping, i.e., larger overlap $\rightarrow$ larger $P$.  This has been shown to have an effect on the \textit{irreversibility} of PolyProtected templates \cite{kh22, s25}, with smaller overlaps leading to more irreversible protected templates; however, \cite{khm26} demonstrated that this difference in irreversibility across different overlaps can be mitigated to a large extent by employing the proposed \textit{key selection algorithm} (see Section \ref{subsec:ks_alg}).

For the PolyProtect evaluations presented in the next sections, we chose $m = 5$, as in \cite{kh22}, meaning that each element in $P$ was generated using 5 consecutive elements from $V$, as illustrated for different overlaps in Fig. \ref{fig:polyprotect}.  As explained in \cite{kh22}, this setting was inspired by the Abel-Ruffini theorem, which states that there is no closed-form algebraic expression for solving polynomials of degree 5 or higher with arbitrary coefficients.  Furthermore, as in \cite{kh22}, we also chose not to set $m > 5$, since this would require using exponents larger than 5 in the PolyProtect transform, which may obliterate small embedding elements.  Consequently, the exponents, $E$, were randomly generated, unique integers in the range [1, 5].  Regarding the choice of values for $C$, we used the [-50, 50] range as in \cite{kh22}, so all sets of $C$s consisted of 5 randomly generated, unique, non-zero integers in this range.

\subsection{Key selection algorithm}
\label{subsec:ks_alg}

In the original PolyProtect work \cite{kh22}, the coefficients, $C$, and exponents, $E$, defining the transformation polynomials were \textit{randomly} generated for each subject (identity).  However, in a recent paper investigating the irreversibility of PolyProtect \cite{khm26}, it was shown that generating $C$ and $E$ randomly may result in protected templates that are \textit{invertible}, in the worst-case scenario where a fully-informed attacker launches an inversion attack using a cosine-distance-based numerical solver (specifically, Python's \textit{scipy.optimize.minimize} function with the \textit{L-BFGS-B} method). So, a \textit{key selection algorithm} was proposed, which helps to ensure that the chosen ``keys'' (i.e., $C$ and $E$ parameters) produce PolyProtected templates that are significantly more difficult (practically impossible) to invert.  This algorithm works as follows.  Assume we have a database of reference templates (e.g., face embeddings) that we wish to protect.  The goal of the key selection algorithm is to choose $C$ and $E$ for those templates, such that when they are transformed via PolyProtect, the corresponding protected templates are irreversible under a numerical solver (in particular, the cosine-distance-based solver).  So, for a given reference template, the algorithm starts by generating $C =  [c_1, c_2, ..., c_5]$ and $E = [e_1, e_2, ..., e_5]$ randomly.  Then, these parameters are used to transform the template to its PolyProtected counterpart, and an inversion attack using the cosine solver is launched.  If the inversion is successful (i.e., the inverted template is close, in terms of cosine distance, to the original face embedding), new $C$ and $E$ are generated, and the process is repeated until the inversion \textit{fails}.  The threshold used to define a successful inversion is set to a value beyond anything likely to be used in a practical face recognition system  (e.g., at 20\% FMR), because if the inversion fails for such a loose threshold, it would be expected to fail for all stricter (more practical) thresholds (e.g., at 0.1\% FMR or 0.01\% FMR).

This key selection algorithm was initially proposed and evaluated for \textit{subject}-specific keys.  In other words, the key selection process was performed separately for each subject enrolled in the face recognition system, to find $C$ and $E$ parameters that work for that particular subject's reference face embedding.  This subject-specific approach makes sense, since PolyProtect was conceived \cite{kh22} to operate in precisely this way, i.e., defining a unique polynomial transformation for each subject.  However, since the ultimate aim of our work is to explore the feasibility of performing identity de-duplication in the \textit{protected} domain, subject-specific transformation parameters are \textit{not suitable}.  This is due to the \textit{unlinkability} property that any BTP method is expected to satisfy (and which PolyProtect, in particular, has already been shown to possess \cite{kh22, s25}).  The unlinkability property stipulates that it should be possible to generate multiple instances of a protected template from the same unprotected template (or, more broadly, the same biometric identity), for example by using different transformation parameters or keys, such that these different protected templates are \textit{not linkable} to each other.  So, in a biometric system protected with a \textit{subject}-specific BTP algorithm, like PolyProtect, a subject could easily bypass a duplication check by enrolling multiple times with the same biometric identity (e.g., face) but with \textit{different protected templates} (generated using different $C$ and $E$ parameters).  In other words, if a BTP method satisfies the \textit{unlinkability} property, then the use of \textit{subject}-specific transformation parameters would make it impossible to perform de-duplication in the protected domain.  For this reason, in our work we explore the use of \textit{system}-specific keys instead of subject-specific keys.  

In a \textit{system}-specific key scenario, \textit{all} subjects enrolled in the protected face recognition system would use the \textit{same} transformation parameters -- in the case of PolyProtect, this means that the \textit{same} $C$ and $E$ parameters would be used to define every subject's transformation polynomial.  This is different from a \textit{subject}-specific scenario, where these parameters are unique for each subject.  When we use system-specific keys, this ensures that the \textit{same transformation} is applied to each subject's face embedding, making it impossible for any one subject to create multiple versions of a protected template by using different $C$ and $E$.  Consequently, performing a duplication check in the protected domain becomes feasible.  However, since \cite{khm26} demonstrated the potential danger of selecting $C$ and $E$ randomly (in terms of accidentally generating an invertible template), and since the proposed key selection algorithm was intended to be used in a subject-specific scenario, our main concern is that it would be extremely difficult to select system-specific $C$ and $E$ parameters that would work well for all subjects (i.e., which would ensure that the generated PolyProtected templates are \textit{irreversible} for everyone).  So, this is the \textbf{focus} of our paper.  In particular, we seek to answer the following question: Is the key selection algorithm from \cite{khm26} generalizable to the \textit{system}-specific key scenario, such that we can select a single key that would generate irreversible protected templates for all subjects (or does it work purely for subject-specific key selection)?  Therefore, is it possible for PolyProtect to safely operate in a \textit{system}-specific key scenario, such that the \textit{irreversibility} BTP criterion is still satisfied?  Section \ref{sec:experiments} presents three experiments to answer this question.

\section{Generating Irreversible Protected Templates using System-Specific Keys}
\label{sec:experiments}

This section presents the main contributions of this paper, in terms of three experiments that investigate the generalizability of the PolyProtect key selection algorithm from \cite{khm26} to the \textit{system}-specific key scenario (towards our ultimate goal of de-duplication in the protected domain).  Since the capability of the key selection algorithm to generate \textit{irreversible} PolyProtected templates has thus far only been demonstrated in the \textit{subject}-specific key scenario \cite{khm26}, the evaluations presented in this section focus on analysing the \textit{irreversibility} of protected templates produced by selecting \textit{system}-specific PolyProtect keys ($C$ and $E$ parameters).  The crux of this analysis is in Sections \ref{subsec:subj_spec} -- \ref{subsec:c_e}, which present the three experiments analysing the applicability of the key selection algorithm \cite{khm26} to system-specific keys: Section \ref{subsec:subj_spec} explores whether it is possible to use keys selected for specific subjects to generate irreversible PolyProtected templates for different subjects, Section \ref{subsec:rand_vec} investigates the use of random vectors (instead of face embeddings) to select system-specific keys that can produce irreversible PolyProtected templates for all subjects' face embeddings, and Section \ref{subsec:c_e} examines the possibility of selecting suitable system-specific keys directly from a ``good'' $C$/$E$ pair distribution (originating from the random-vector-based keys in Section \ref{subsec:rand_vec}).  These three sections are preceded by Section \ref{subsec:set_up}, which details the experimental set-up (i.e., the face recognition models selected for the extraction of face embeddings, to which PolyProtect is applied, and the face datasets serving as the image sources for these embeddings), and Section \ref{subsec:motivation}, which motivates our focus on system-specific keys by empirically demonstrating the unsuitability of subject-specific keys for de-duplication in the protected domain.  Section \ref{subsec:accuracy} concludes our experiments on system-specific keys by investigating the effect of selected keys on the recognition accuracy of the protected face recognition system.

\subsection{Set-up: Face recognition models and datasets}
\label{subsec:set_up}

In order to evaluate PolyProtect in the system-specific key scenario, within the context of work focusing on the protection of face data, the first step was to generate the face templates to which PolyProtect would be applied.  Face recognition systems commonly employ neural-network-based models for the extraction of face ``embeddings'' (templates), so these were the types of feature extractors we were interested in.  To select the best face recognition model(s) for our study, we evaluated five state-of-the-art models to which we have access: iResNet50 and iResNet100\footnote{\url{https://github.com/deepinsight/insightface/tree/master/model_zoo}} \cite{h16, dg19}, EdgeFace and EdgeFace-XS\footnote{\url{https://github.com/otroshi/edgeface}} \cite{g24}, and FaceNet\footnote{\url{https://github.com/timesler/facenet-pytorch}} \cite{s15}.  These models were applied to three face datasets, to extract the (512-dimensional) face embeddings from the underlying face images.  The datasets used in our work were the same as those employed in \cite{khm26}, which were selected to represent three different face image acquisition environments:

\begin{itemize}
	\item Multi-PIE\footnote{\url{https://www.cs.cmu.edu/afs/cs/project/PIE/MultiPie}} \cite{g10}:  Contains face images of 337 subjects.  The images were captured in a very controlled environment, using multiple cameras fixed at different angles.  We randomly selected 10 images per subject across the 3 frontal cameras (14\_0, 05\_1, and 05\_0), which resulted in a total of 3,370 face images.
	\item SOTERIA\footnote{\url{https://www.idiap.ch/en/scientific-research/data/soteria}} \cite{r24}: Contains face videos of 70 subjects.  The (bona-fide) videos were captured in a less controlled environment than Multi-PIE, using the frontal (``selfie'') and back cameras of five mobile phones (Apple iPhones 6s and 12, Xiaomi Redmi 6 Pro and 9A, and Samsung Galaxy S9), under various lighting conditions.  We randomly selected 10 frontal frames per subject across the five phones, which resulted in a total of 700 face images.
	\item iCarB-Face\footnote{\url{https://www.idiap.ch/en/scientific-research/data/icarb-face}} \cite{kh24}: Contains face videos of 197 subjects.  The videos were captured inside a car, using an NIR camera, while the subjects were seated in the driver's seat.  We selected 4 video frames per subject, with a neutral facial expression and no accessories: 2 when the car was parked indoors and 2 when it was parked outdoors, with the subject's eyes open in one image and closed in the other.  This resulted in a total of 788 face images.
\end{itemize}  

\begin{figure}[!h]
\centering
\subfloat{\includegraphics[width=0.33\columnwidth]{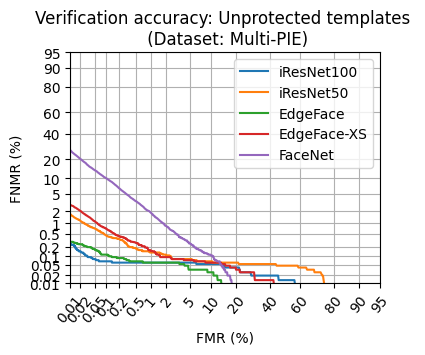}}
\hfil
\subfloat{\includegraphics[width=0.33\columnwidth]{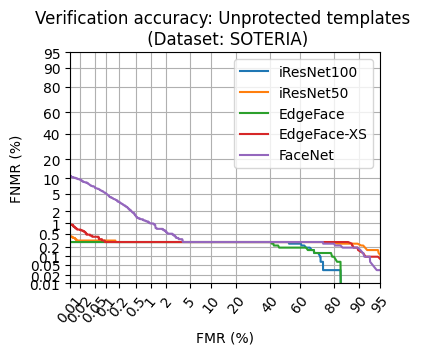}}
\hfil
\subfloat{\includegraphics[width=0.33\columnwidth]{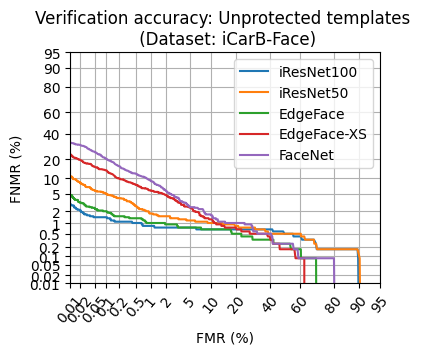}} 
\caption{Verification accuracy across unprotected face templates (embeddings) generated using five different face recognition models on three face datasets.}
\label{fig:verification}
\end{figure}

To select the best face recognition model(s) for our study, the verification accuracy for each of the five sets of extracted face embeddings was computed on the three datasets.  Within each set, the score (cosine distance) was calculated between all pairs of embeddings.  Fig. \ref{fig:verification} compares the resulting accuracy across the five face recognition models in terms of the False Non-Match Rate (FNMR) and False Match Rate (FMR).

From Fig. \ref{fig:verification}, we see that EdgeFace and iResNet100 achieve the highest accuracy and are very close in their performance.  So, the experiments presented in Sections \ref{subsec:motivation} -- \ref{subsec:accuracy} are based on face embeddings extracted using these two models.

\textit{Notes:} All upcoming experimental results are reported for each of the two selected face recognition models (EdgeFace and iResNet100) and three aforementioned datasets; however, due to space constraints, we present our findings only for the ``overlap = 3'' PolyProtect configuration (see Fig. \ref{fig:polyprotect}), since this was the setting recommended for best irreversibility/accuracy trade-off in \cite{khm26}.  Similar observations (and the same global conclusions) were noted across all overlaps -- the interested reader can confirm this by reproducing the experiments using our open-source code.  Furthermore, as per the recommendation in \cite{khm26}, the PolyProtected templates used for all experiments were created from \textit{normalized} face embeddings.

\subsection{Why use system-specific keys?}
\label{subsec:motivation}

Before launching into the system-specific key selection methods investigated for the PolyProtect BTP method in our work (Sections \ref{subsec:subj_spec} -- \ref{subsec:c_e}), we begin with an illustration of \textit{why} system-specific key selection is even necessary.  In Sections \ref{sec:introduction} and \ref{subsec:ks_alg}, we motivated our interest in \textit{system}-specific keys by explaining that \textit{subject}-specific keys would prevent us from performing de-duplication of face identities in the PolyProtected template domain (the ultimate aim of this work), due to the clash with the \textit{unlinkability} property of this BTP method.  In particular, it was noted that the use of subject-specific keys could enable a person to enroll multiple times into the same face recognition system, using different keys ($C$ and $E$ parameters for the PolyProtect transform) to generate different (non-matching) protected templates.  In this section, we empirically validate this claim.

To understand the problem with using \textit{subject}-specific PolyProtect keys, and thus the advantage of \textit{system}-specific keys, when aiming to perform de-duplication in the protected domain, we conducted the following experiment.  Using the face embeddings from the EdgeFace and iResNet100 models and our three face datasets, we generated PolyProtected templates using: (i) randomly generated \textit{system}-specific keys (i.e., the \textit{same} $C$ and $E$ parameters were applied to \textit{all} the face embeddings in the same dataset, across all identities), and (ii) randomly generated \textit{sample}-specific keys (i.e., \textit{different} $C$ and $E$ parameters were applied to \textit{each face embedding}, including different embeddings from the same face identity).  The second scenario represents the potential issue with allowing the use of subject-specific keys, whereby the same person could apply different $C$ and $E$ parameters to (a different instance of) the face embedding representing the same identity, to generate a \textit{different} (non-matching, unlinkable) PolyProtected template.  

\begin{figure}[!h]
\centering
\subfloat{\includegraphics[width=0.3\columnwidth]{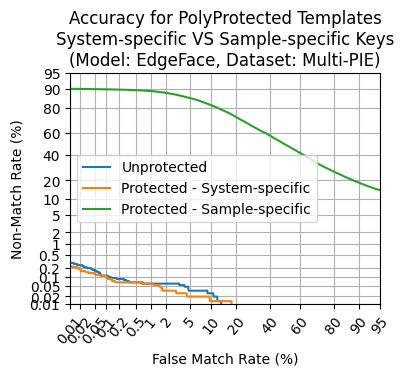}}
\hfil 
\subfloat{\includegraphics[width=0.3\columnwidth]{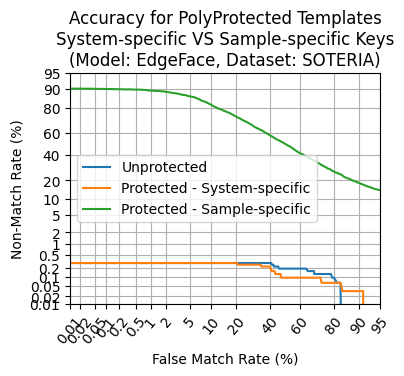}}
\hfil
\subfloat{\includegraphics[width=0.3\columnwidth]{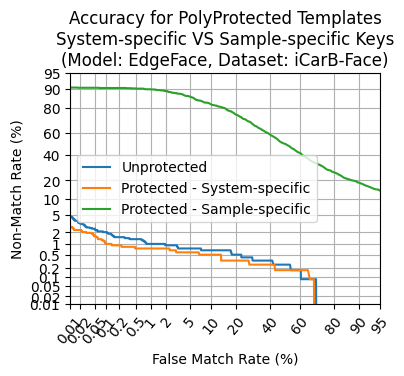}}
\vfil
\subfloat{\includegraphics[width=0.3\columnwidth]{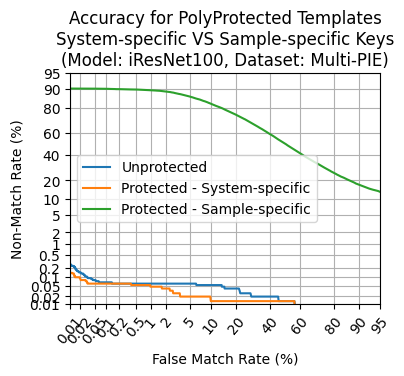}}
\hfil 
\subfloat{\includegraphics[width=0.3\columnwidth]{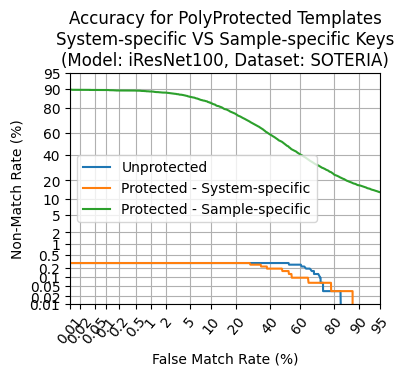}}
\hfil
\subfloat{\includegraphics[width=0.3\columnwidth]{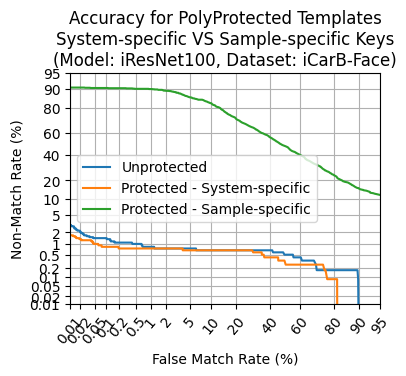}}
\caption{Accuracy of template matching in the PolyProtected domain, for protected templates created from EdgeFace and iResNet100 face embeddings in two scenarios: (i) \textit{System}-specific keys used for PolyProtect (i.e., the \textit{same} $C$ and $E$ parameters used for all face embeddings across all identities), and (ii) \textit{Sample}-specific keys used for PolyProtect (i.e., \textit{different} $C$ and $E$ parameters used for each face embedding, including different embeddings from the same identity).  Accuracy for unprotected embeddings is shown as the baseline.}
\label{fig:motivation}
\end{figure}

Fig. \ref{fig:motivation} shows the accuracy in the PolyProtected domain in these two scenarios, in terms of the Non-Match Rate (percentage of protected templates from the \textit{same identity} that do \textit{not} match) versus the False Match Rate (percentage of protected templates from \textit{different identities} that \textit{do} match).  

From Fig. \ref{fig:motivation} it is clear that the Non-Match Rate in the \textit{sample}-specific key scenario is extremely high: around 90\% at FMR of 0.01\%, 0.1\%, and 1\%.  This suggests that, if a person chose to enroll in the same face recognition system multiple times with the same face identity, using different PolyProtect keys ($C$ and $E$ parameters) to protect each face embedding, 90\% of the time the resulting protected templates would be flagged as belonging to \textit{different identities}, i.e., de-duplication would \textit{fail} 90\% of the time.  This supports our earlier claim that de-duplication of face identities in the protected domain is practically \textit{impossible} when \textit{subject}-specific keys are used for the PolyProtect transform, since this opens up the possibility for multiple enrollments of the same identity with different PolyProtect keys, which would \textit{not} be flagged as the same identity during de-duplication (due to PolyProtect's \textit{unlinkability} property).  On the other hand, we see that when \textit{system}-specific PolyProtect keys are used, the Non-Match Rate in the protected domain is significantly lower, approximately the same as the accuracy in the unprotected domain.  This tells us that successful detection of multiple enrollments of the same identity in the protected domain when \textit{system}-specific keys are used to transform the face embeddings to their protected counterparts, is about as likely as successful de-duplication on \textit{unprotected} face templates.  So, our claim that de-duplication in the protected domain \textit{is} possible when PolyProtect is based on \textit{system}-specific keys, is confirmed.  This motivates our investigation into the system-specific key selection methods presented in Sections \ref{subsec:subj_spec} -- \ref{subsec:c_e}.

\subsection{System-specific key selection \#1: From subject-specific keys}
\label{subsec:subj_spec}

This section presents the first experiment conducted to investigate the generalizability of the PolyProtect key selection algorithm from \cite{khm26} to the \textit{system}-specific key scenario.  The aim was to see whether a ``key'' (i.e., a set of $C$ and $E$ parameters) selected for a particular subject's face embedding could be applied to a \textit{different} subject's face embedding to still generate an \textit{irreversible} protected template.

To perform this experiment, we proceeded as follows.  We used the key selection algorithm from \cite{khm26} to select \textit{subject}-specific keys for all subjects from each of our three face datasets, when the face embeddings were generated from the underlying face images using the two face recognition models (feature extractors) chosen in Section \ref{subsec:set_up}: EdgeFace and iResNet100.  Then, 10 subjects' keys (i.e., 10 sets of $C$ and $E$ parameters) were selected at random and applied (one at a time) to the face embeddings of \textit{all} the subjects.  In other words, the experiment consisted of 10 trials corresponding to 10 different system-specific keys, separately for EdgeFace and iResNet100 embeddings from each of our three datasets.  The worst-case, cosine-solver-based inversion attack from \cite{khm26} was then launched on the resulting protected templates (created by applying PolyProtect with the selected keys to the face embeddings) to see if they could be inverted -- if \textit{not}, this would indicate that the selected keys were good enough to produce \textit{irreversible} protected templates.  

\begin{figure*}[!h]
\centering
\subfloat{\includegraphics[width=0.3\textwidth]{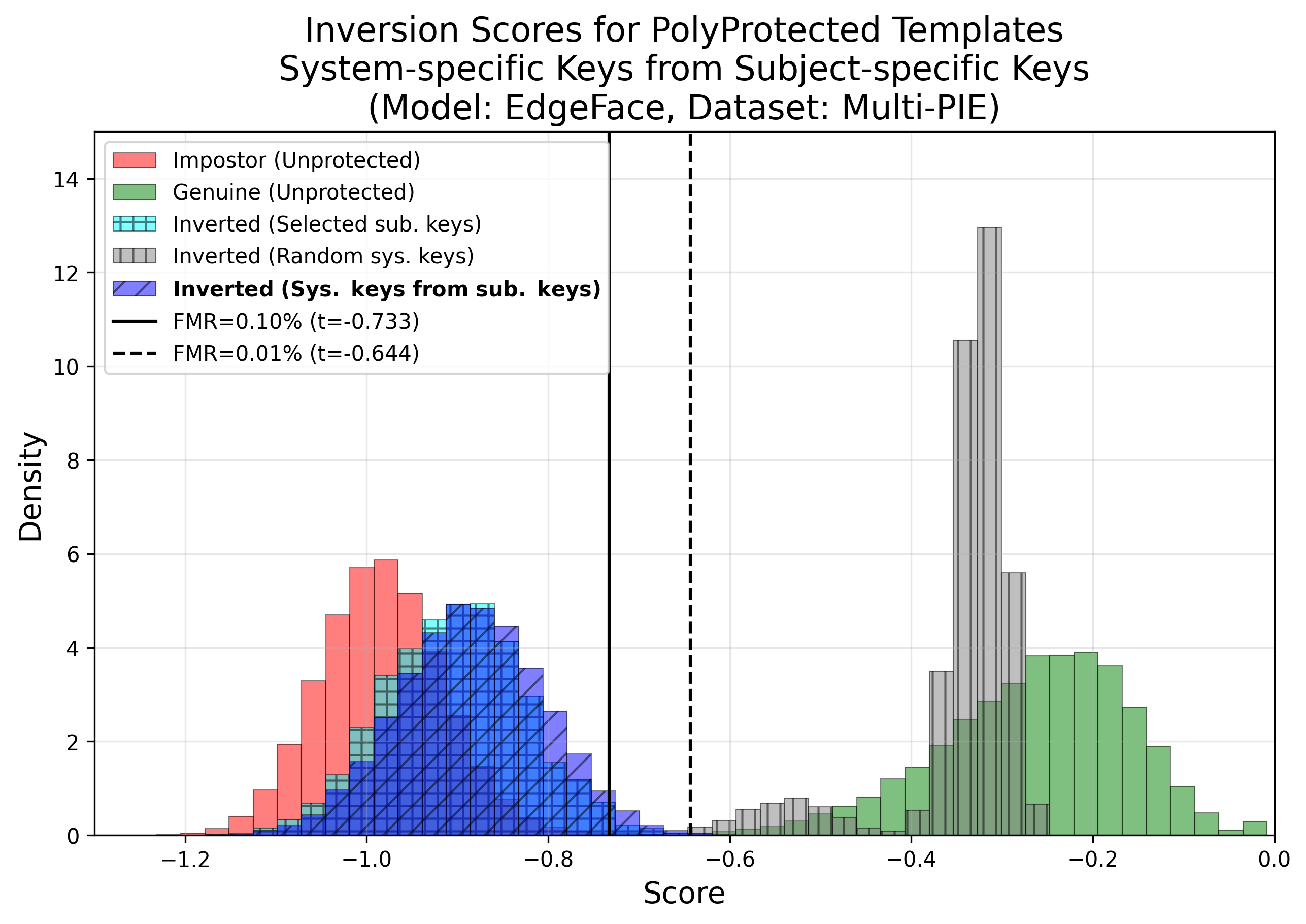}}
\hfil 
\subfloat{\includegraphics[width=0.3\textwidth]{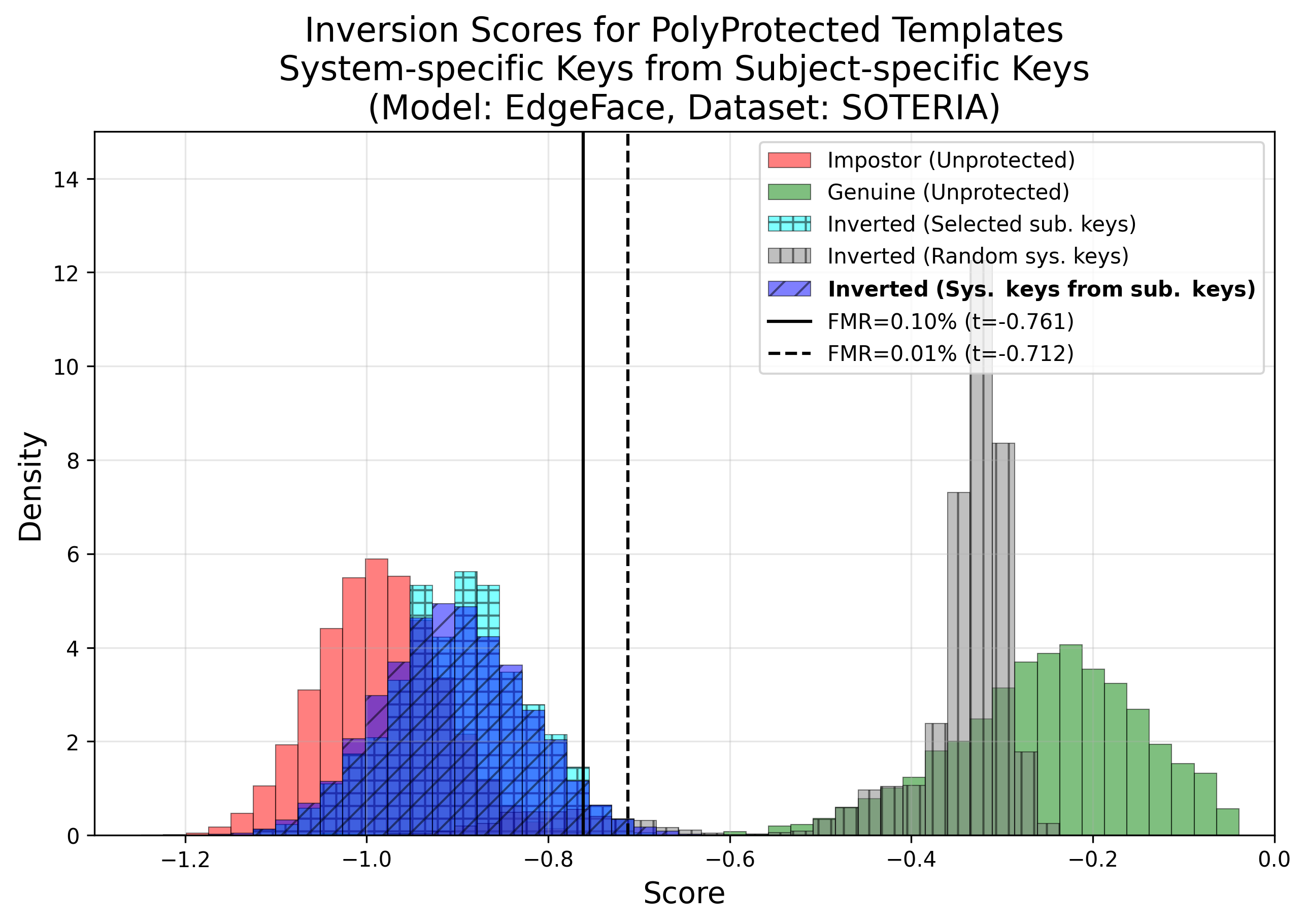}}
\hfil
\subfloat{\includegraphics[width=0.3\textwidth]{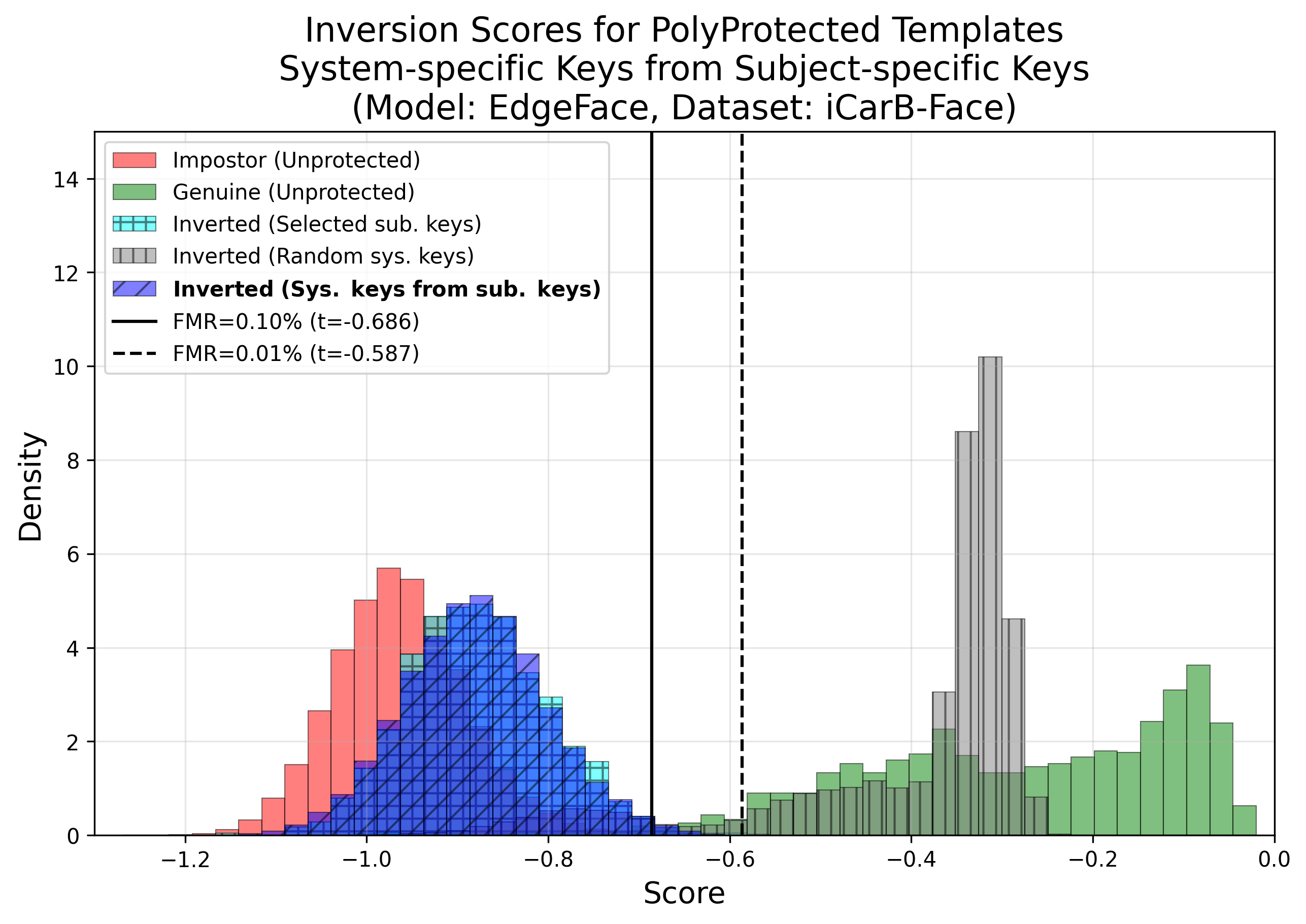}}
\vfil
\subfloat{\includegraphics[width=0.3\textwidth]{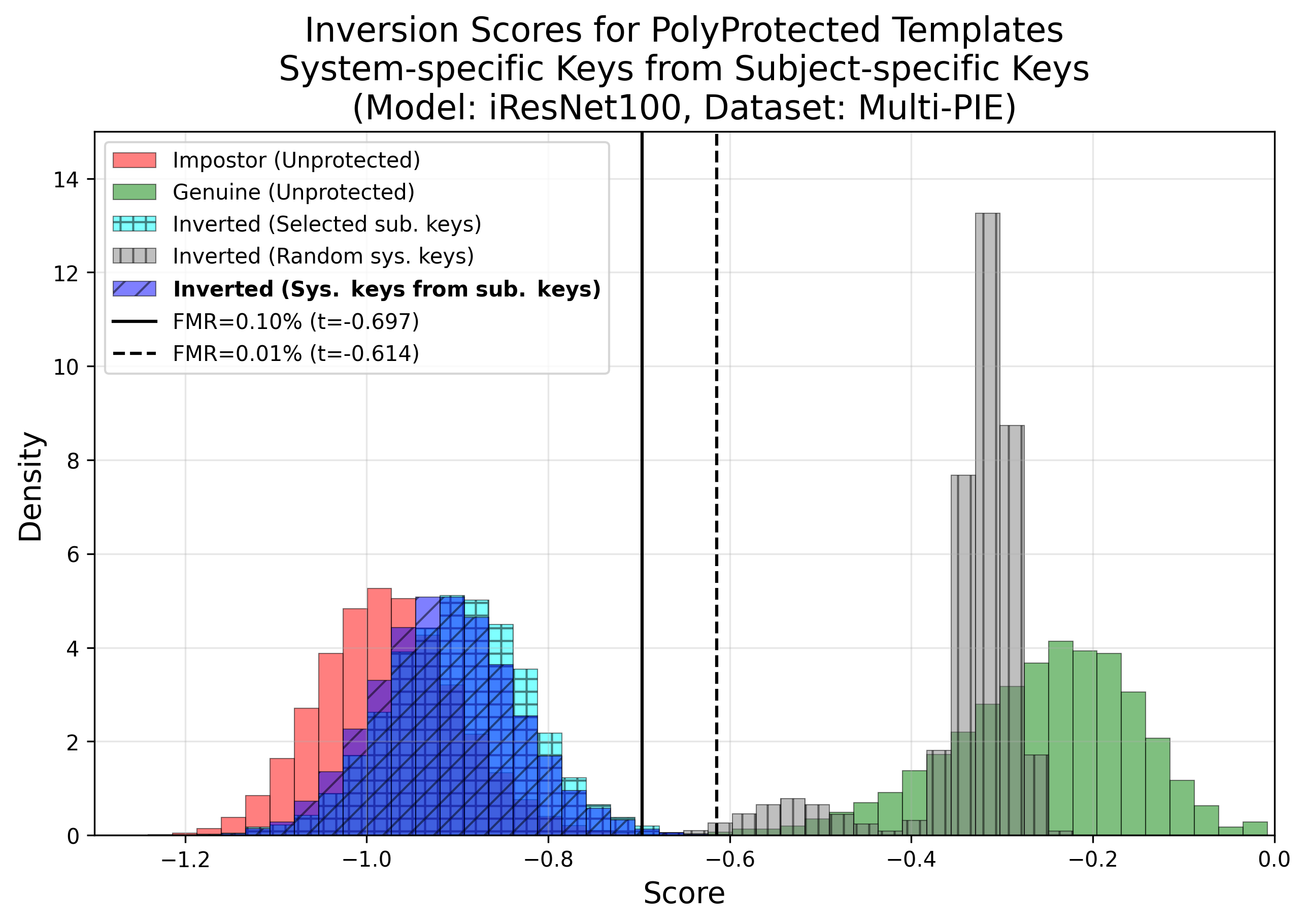}}
\hfil
\subfloat{\includegraphics[width=0.3\textwidth]{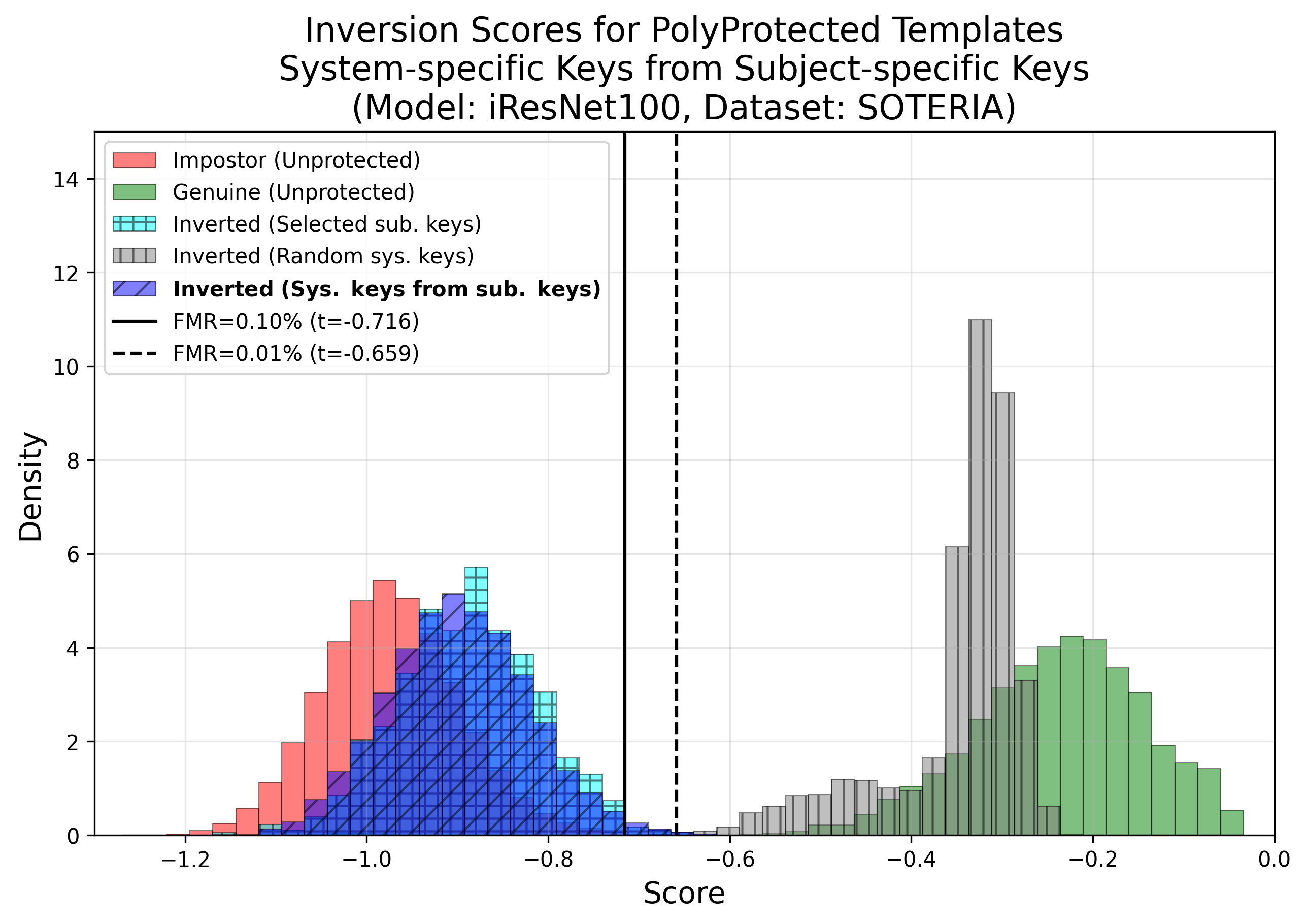}}
\hfil
\subfloat{\includegraphics[width=0.3\textwidth]{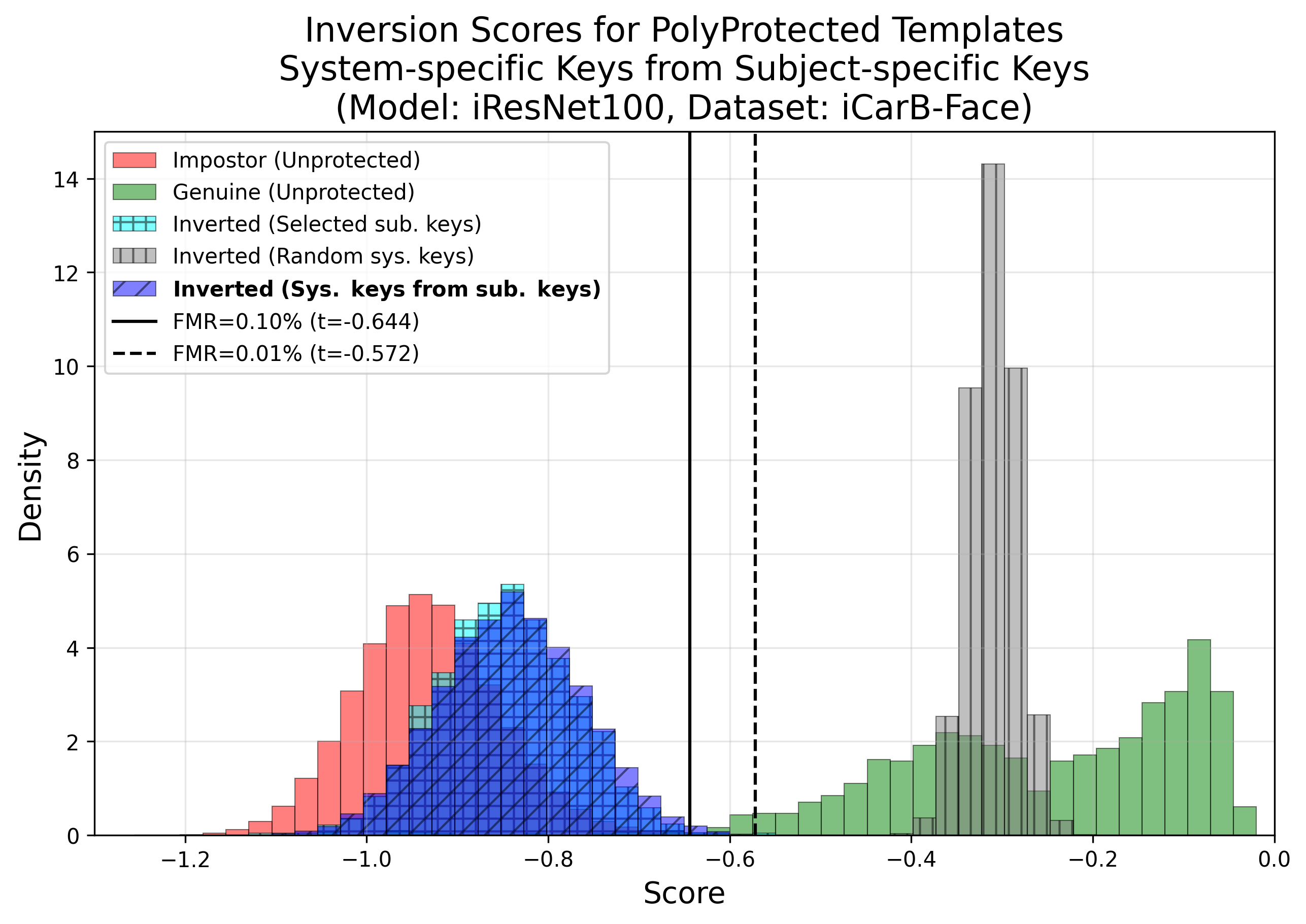}}
\caption{Inversion scores for PolyProtected templates generated from EdgeFace and iResNet100 face embeddings when the keys ($C$, $E$) were: system-specific + random (gray histogram), subject-specific + selected using the key selection algorithm from \cite{khm26} (cyan histogram), and system-specific + selected at random from the subject-specific keys (blue histogram).  Our main focus is the \textbf{blue} histogram.  The inversion was performed using the cosine-based numerical solver from \cite{khm26}.  The two vertical black lines represent different thresholds at which the inversion success rate (ISR) was computed for Table \ref{tab:expt_1}.}
\label{fig:expt_1}
\end{figure*}

Fig. \ref{fig:expt_1} shows histograms representing the inversion scores (i.e., cosine distances\footnote{Multiplied by -1 to turn them into \textit{similarity} scores.}) between the inverted templates and the true face embeddings from which the protected templates were originally created.  The gray histogram represents protected templates generated using 10 \textit{random} system-specific keys, in which case the key selection algorithm from \cite{khm26} was \textit{not} used.  Then, the cyan histogram corresponds to protected templates generated using \textit{subject}-specific keys chosen using the key selection algorithm, and the blue histogram contains inversion scores for protected templates generated using 10 \textit{system}-specific keys selected (at random) from these subject-specific keys -- our main focus is this last histogram.  All histograms were produced by launching an inversion attack on each protected template 10 times, using 10 different initial guesses for the numerical solver (Python's \textit{scipy.optimize.minimize} function with the \textit{L-BFGS-B} method), and concatenating the resulting inversion scores.  Each inversion score indicates how close, in terms of cosine distance, the inverted template is to the original face embedding. Fig. \ref{fig:expt_1} also shows histograms for the genuine and impostor scores computed on the unprotected face embeddings.  The inversion scores should lie as close as possible to the impostor distribution, which would indicate that the inverted templates are \textit{as different} from their corresponding face embeddings as are embeddings from different identities.  

Table \ref{tab:expt_1} quantifies the Fig. \ref{fig:expt_1} results in terms of Inversion Success Rate (ISR).  As in \cite{kh22, khm26}, the ISR represents the proportion of PolyProtected templates whose inversion score is below a pre-defined threshold, in which case the inversion attack would be considered successful.  We present results at the two thresholds illustrated in Fig. \ref{fig:expt_1}: at 0.1\% and 0.01\% FMR, computed on the unprotected face recognition system (genuine and impostor scores).  

\begin{table}[!h]
\renewcommand{\arraystretch}{1.2}
\caption{Inversion success rate (ISR) for Fig. \ref{fig:expt_1} at the two FMR thresholds, for EdgeFace and iResNet100 face embeddings protected via PolyProtect when the keys ($C$, $E$) were: subject-specific + selected using the key selection algorithm from \cite{khm26} (\textit{Sb.}) VS \textbf{system-specific + selected from the subject-specific keys (\textit{Sy.})} VS system-specific + random (\textit{R-Sy.}). \label{tab:expt_1}} 
\centering
\begin{tabular}{|c|c|c|c|c|c|c|c|}
\hline
\multirow{3}{*}{Model} & \multirow{3}{*}{Dataset} & \multicolumn{6}{c|}{ISR (\%)} \\
\cline{3-8}
 & & \multicolumn{3}{c|}{@ 0.1\% FMR} & \multicolumn{3}{c|}{@ 0.01\% FMR} \\
\cline{3-8}
 & & \textit{Sb.} & \textbf{\textit{Sy.}} & \textit{R-Sy.} & \textit{Sb.} & \textbf{\textit{Sy.}} & \textit{R-Sy.} \\
\hline
\multirow{3}{*}{EdgeFace} & Multi-PIE & 0.5 & \textbf{0.8} & 100 & 0 & \textbf{0} & 99.2 \\
\cline{2-8}
 & SOTERIA & 0 & \textbf{0.6} & 91.3 & 0 & \textbf{0} & 90.1 \\
\cline{2-8}
 & iCarB-Face & 1.1 & \textbf{1.3} & 91.6 & 0.2 & \textbf{0} & 89.7 \\
\hline
\multirow{3}{*}{iResNet100} & Multi-PIE & 0.5 & \textbf{0.5} & 100 & 0 & \textbf{0} & 99.4 \\
\cline{2-8}
 & SOTERIA & 0.6 & \textbf{0.7} & 100 & 0 & \textbf{0} & 99.6 \\
\cline{2-8}
 & iCarB-Face & 2.2 & \textbf{3.4} & 100 & 0.2 & \textbf{0.2} & 100 \\
\hline
\end{tabular}
\end{table} 

In Fig. \ref{fig:expt_1} we observe that the gray histogram is located further to the right (overlapping with the green unprotected template \textit{genuine} score distribution) than the cyan and blue histograms (overlapping with the red unprotected template \textit{impostor} score distribution).  This tells us that choosing system-specific keys ($C$, $E$) \textit{randomly} is more likely to produce \textit{invertible} PolyProtected templates, compared to when these keys are chosen using the key selection algorithm from \cite{khm26} (represented by the cyan and blue histograms).  Considering Table \ref{tab:expt_1}, we see that this is indeed the case: the ISR for both face recognition models, across all three datasets, is significantly lower in the scenario where the keys are selected using the key selection algorithm from \cite{khm26} and are either subject-specific (\textit{Sb.}) or else system-specific and chosen from the set of pre-generated subject-specific keys (\textit{Sy.}), compared to when the keys are system-specific and generated in a purely random fashion (\textit{R-Sy.}).  This confirms the observation and recommendation from \cite{khm26}, which was that purely random selection of $C$ and $E$ should be avoided in favour of using the key selection algorithm; our \textbf{novel finding} is that this is true in the \textit{system}-specific key selection scenario too, and not only the subject-specific scenario explored in \cite{khm26}.  

Another important \textbf{new finding} in our work, which is clear from the overlap between the cyan and blue histograms in Fig. \ref{fig:expt_1}, is that system-specific keys chosen (at random) from a list of subject-specific keys generated using the key selection algorithm from \cite{khm26}, are approximately as effective as the subject-specific keys at producing \textit{irreversible} PolyProtected templates, i.e., the inversion scores are as low in the system-specific key scenario (blue histogram) as in the subject-specific key scenario (cyan histogram).  Table \ref{tab:expt_1} more or less confirms this observation, although in some cases, especially at the 0.1\% FMR threshold, the ISR for system-specific keys chosen from subject-specific keys (\textit{Sy.}) is slightly higher than the ISR in the subject-specific key scenario (\textit{Sb.}).  This may be due to the fact that the subject-specific keys were selected to work for particular subjects' face embeddings, so it could happen that sometimes these keys do not work equally well for \textit{all} subjects' embeddings in a system-specific key scenario; however, since all the \textit{Sb.} and \textit{Sy.} ISR values, especially at the stricter threshold (0.01\% FMR), are very low, any differences between the two key selection scenarios seem fairly insignificant (e.g., compared to the considerable difference between \textit{Sb.}/\textit{Sy.} and \textit{R-Sy.}).  So, we may conclude that subject-specific keys chosen using the key selection algorithm from \cite{khm26} \textit{generalize} well to the system-specific key scenario.

\subsection{System-specific key selection \#2: From random vectors}
\label{subsec:rand_vec}

Section \ref{subsec:subj_spec} showed that subject-specific PolyProtect keys ($C$, $E$), selected using the key selection algorithm from \cite{khm26}, can be applied as \textit{system}-specific keys to generate \textit{irreversible} PolyProtected templates across an entire database of different subjects' face embeddings.  So, we already know that the key selection algorithm \textit{generalizes} to the \textit{system}-specific key scenario.  In this section, we take the generalizability investigation a step further, by conducting an experiment to see if we can select system-specific keys from (512-dimensional) \textit{random vectors}, instead of from the set of keys chosen for specific subjects' face embeddings (as in Section \ref{subsec:subj_spec}).  The aim was to see whether keys can be selected independently of the face embeddings to which they will be applied -- while this was demonstrated to some extent by the experiment in Section \ref{subsec:subj_spec}, the use of random vectors for key selection in this second experiment aims to completely decouple this process from the structure of the target face embeddings.  

\begin{figure*}[!h]
\centering
\subfloat{\includegraphics[width=0.3\textwidth]{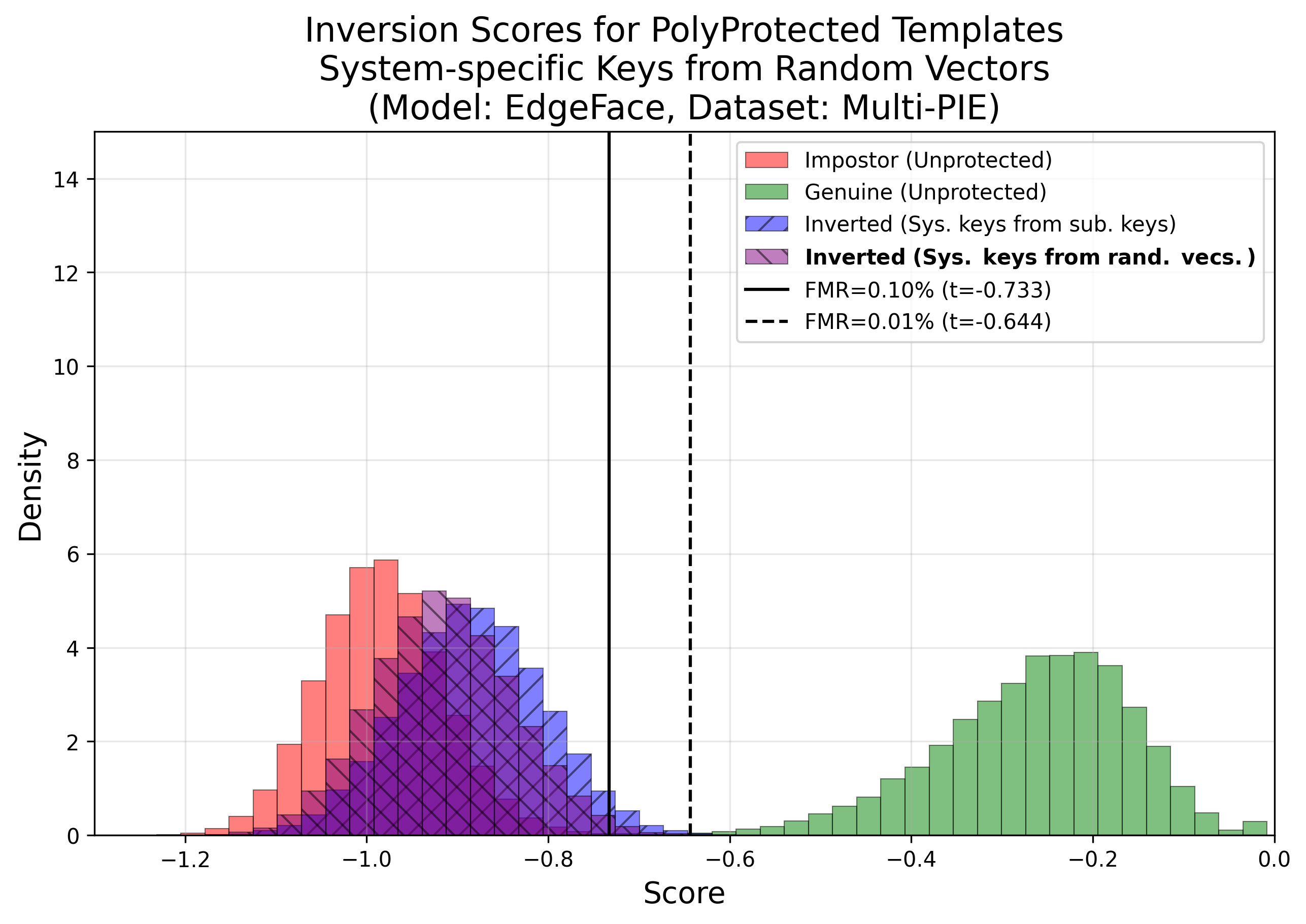}}
\hfil 
\subfloat{\includegraphics[width=0.3\textwidth]{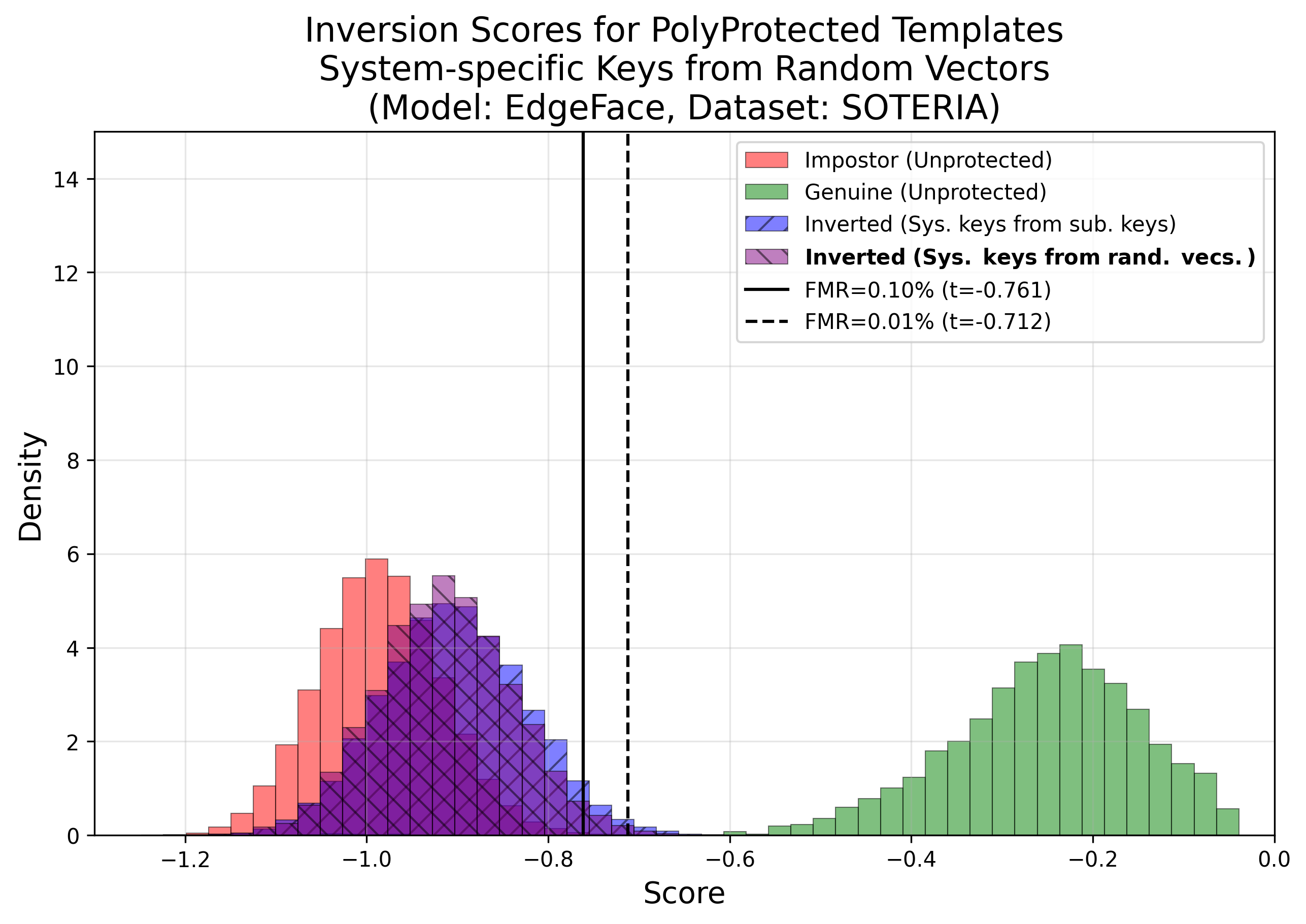}}
\hfil
\subfloat{\includegraphics[width=0.3\textwidth]{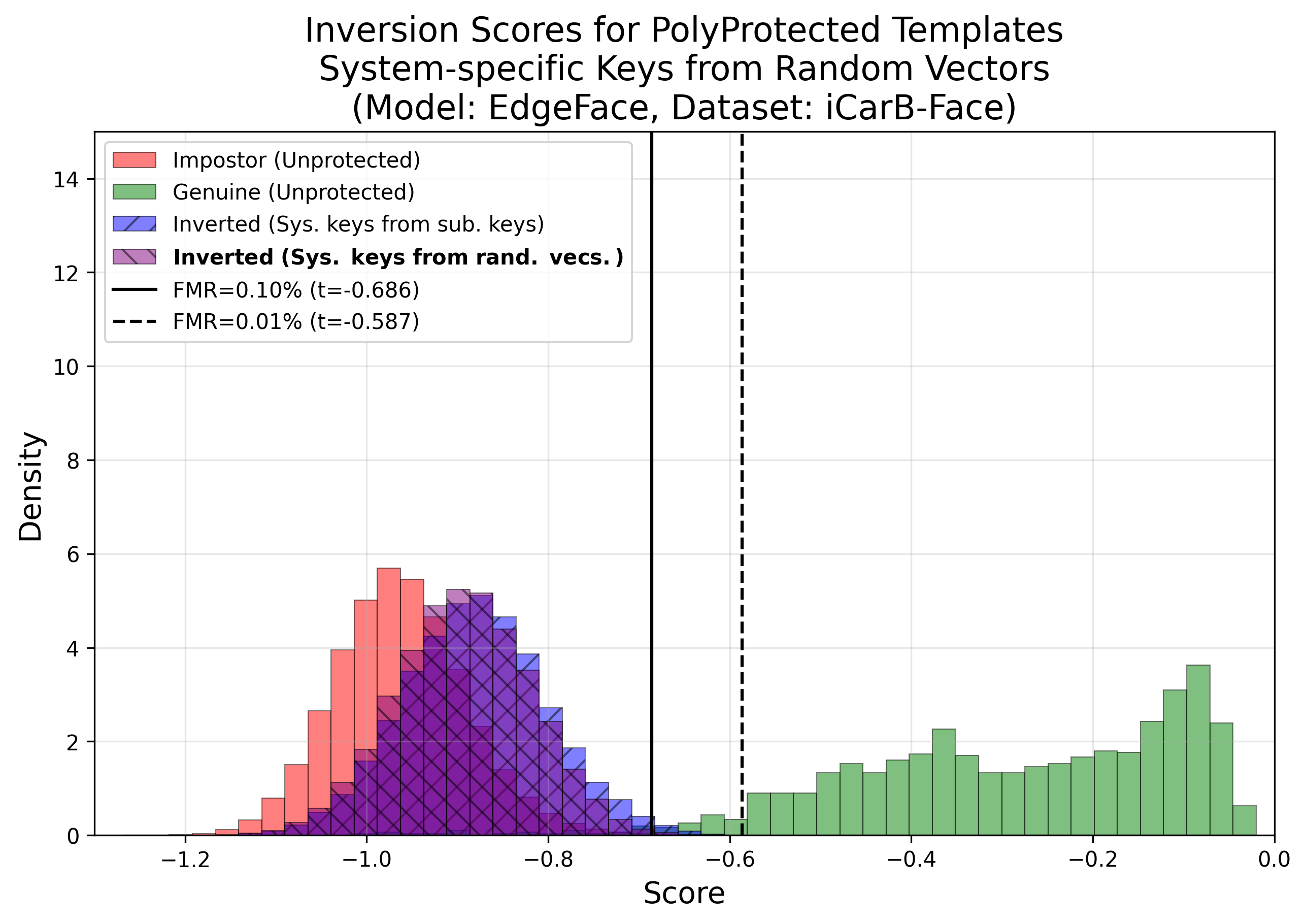}}
\vfil
\subfloat{\includegraphics[width=0.3\textwidth]{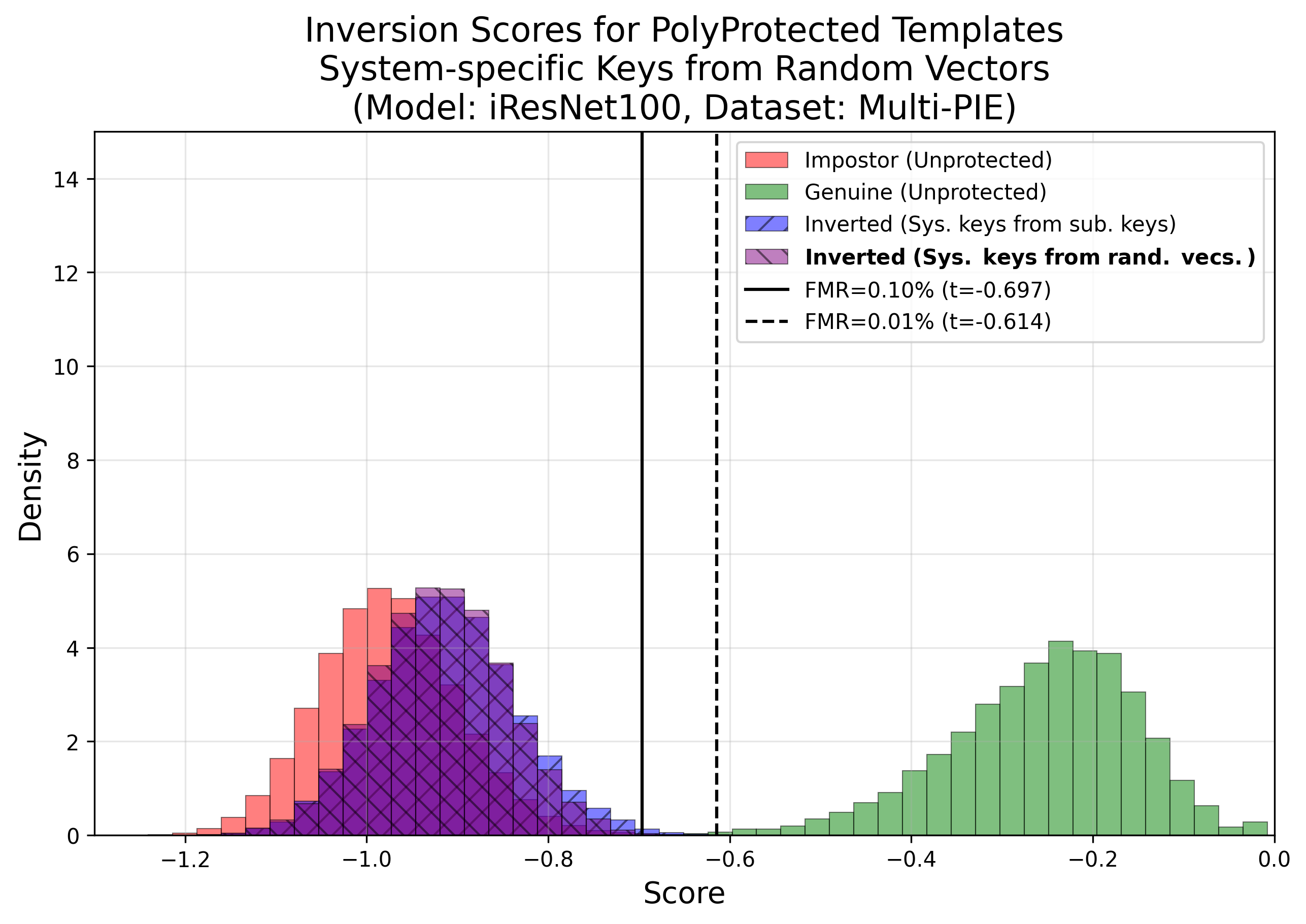}}
\hfil
\subfloat{\includegraphics[width=0.3\textwidth]{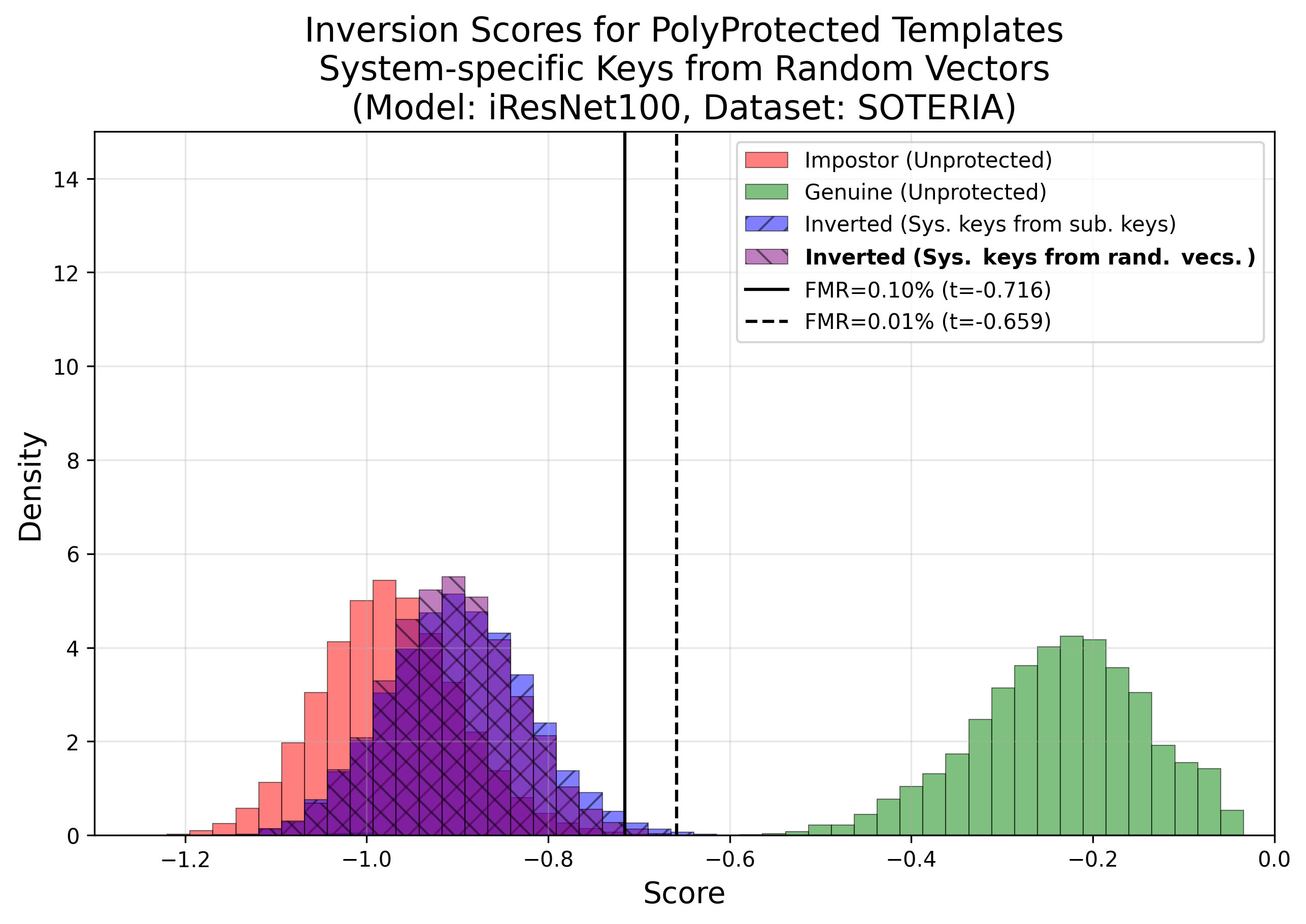}}
\hfil
\subfloat{\includegraphics[width=0.3\textwidth]{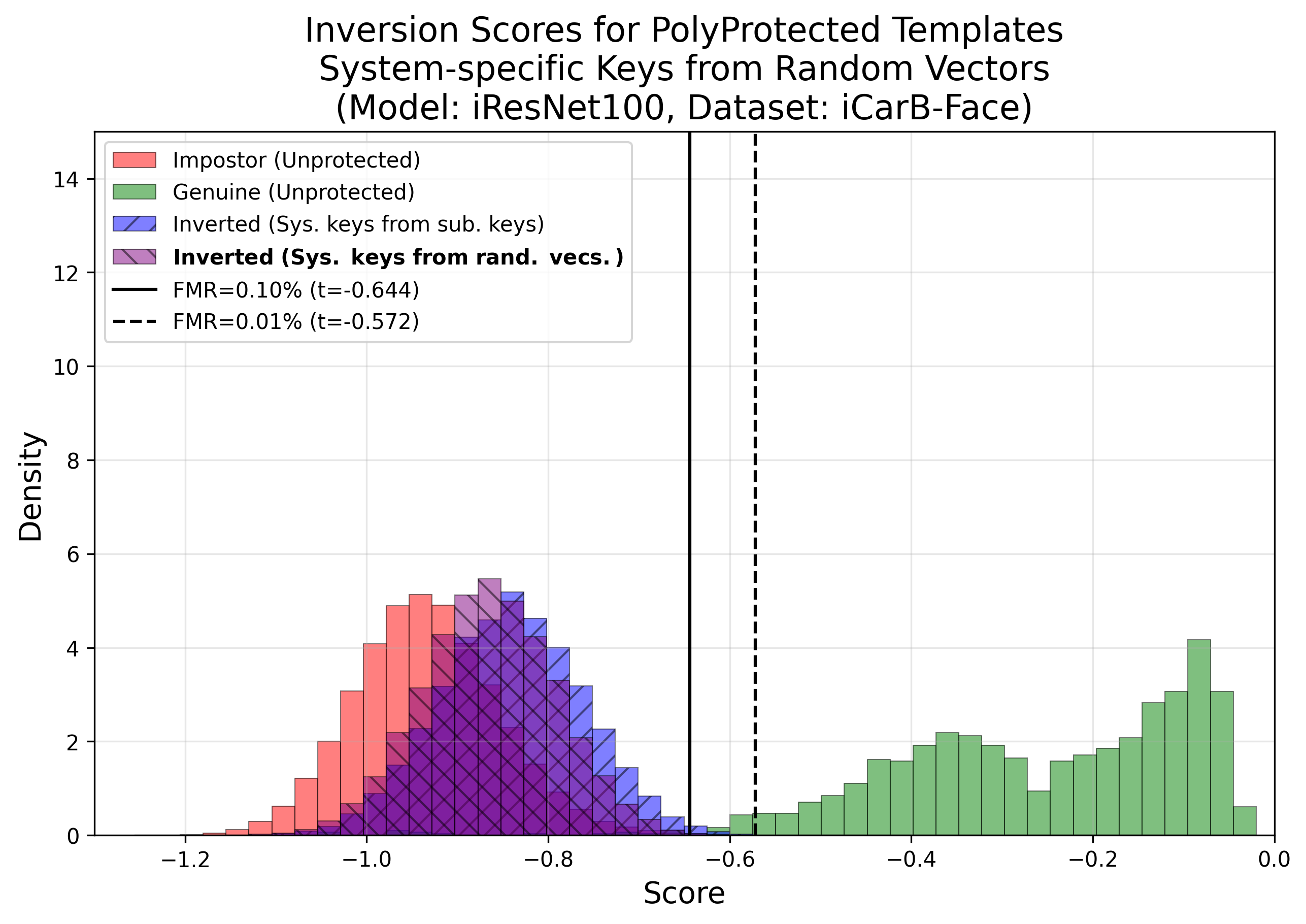}}
\caption{Inversion scores for PolyProtected templates generated from EdgeFace and iResNet100 face embeddings when the keys ($C$, $E$) were system-specific and: selected by applying the key selection algorithm from \cite{khm26} to random vectors (purple histogram) versus chosen at random from subject-specific keys selected by applying the same key selection algorithm to face embeddings (blue histogram, from Fig. \ref{fig:expt_1}).  Our main focus is the \textbf{purple} histogram.  The inversion was performed using the numerical solver from \cite{khm26}.  The two vertical black lines represent different thresholds at which the inversion success rate (ISR) was computed for Table \ref{tab:expt_2}.}
\label{fig:expt_2}
\end{figure*}

To perform this experiment, we proceeded as follows.  Instead of selecting 10 system-specific keys from a pool of subject-specific keys chosen for individual face embeddings, as in Section \ref{subsec:subj_spec}, this time we tried to select the 10 system-specific keys from \textit{random vectors}, whose value range is the same as the range of the (normalized) EdgeFace and iResNet100 face embeddings.  More specifically, we ran the key selection algorithm from \cite{khm26} on random vectors for a total of 48 hours.  A key was considered ``good'' if, when the corresponding $C$ and $E$ parameters were applied to a particular random vector, the resulting PolyProtected template was \textit{irreversible}.  To check for irreversibility, an inversion attack using the numerical solver from \cite{khm26} was launched.  If the solver did \textit{not} manage to invert the protected template to recover a close approximation of the original vector (i.e., the similarity, in terms of negative cosine distance, between the inverted template and the vector was less than a specific threshold\footnote{We used a threshold of -1.0 to force the inversion score histogram to overlap as much as possible with the impostor histogram (based on Fig. \ref{fig:expt_1}).}), the key was considered ``good''.  This process was repeated for as many random vectors as possible within the 48-hour period.  Then, from the entire set of ``good'' keys found in that time, 10 keys were randomly selected to represent \textit{system}-specific keys that would be applied to \textit{all} subjects' face embeddings (a different set of 10 keys was selected for EdgeFace and iResNet100 embeddings from each of our three datasets).  The cosine-solver-based inversion attack was then launched on the protected templates 10 times, using 10 random initial guesses (as in Section \ref{subsec:subj_spec}), to generate inversion scores and check if the protected face templates are irreversible (in terms of ISR).  Fig. \ref{fig:expt_2} compares the resulting (purple) histogram of inversion scores (when face embeddings are protected using system-specific keys generated by applying the key selection algorithm to \textit{random vectors}) to the histogram of inversion scores when system-specific keys come from subject-specific keys selected for specific subjects' face embeddings (i.e., the blue histogram from Fig. \ref{fig:expt_1}).  Table \ref{tab:expt_2} compares the ISRs at the same two FMR thresholds used for Table \ref{tab:expt_1}.

\begin{table}[!h]
\renewcommand{\arraystretch}{1.2}
\caption{Inversion success rate (ISR) for Fig. \ref{fig:expt_2} at the two FMR thresholds, for EdgeFace and iResNet100 face embeddings protected via PolyProtect when the keys ($C$, $E$) were: \textbf{system-specific + selected using the key selection algorithm from \cite{khm26} applied to random vectors (\textit{RV})} VS system-specific + chosen from subject-specific keys selected for individual face embeddings (\textit{SK} -- corresp. to \textit{Sy.} in Table \ref{tab:expt_1}). \label{tab:expt_2}} 
\centering
\begin{tabular}{|c|c|c|c|c|c|}
\hline
\multirow{3}{*}{Model} & \multirow{3}{*}{Dataset} & \multicolumn{4}{c|}{ISR (\%)} \\
\cline{3-6}
 & & \multicolumn{2}{c|}{@ 0.1\% FMR} & \multicolumn{2}{c|}{@ 0.01\% FMR} \\
\cline{3-6}
 & & \textbf{\textit{RV}} & \textit{SK} & \textbf{\textit{RV}} & \textit{SK} \\
\hline
\multirow{3}{*}{EdgeFace} & Multi-PIE & \textbf{0.3} & 0.8 & \textbf{0} & 0 \\
\cline{2-6}
 & SOTERIA & \textbf{0.2} & 0.6 & \textbf{0} & 0 \\
\cline{2-6}
 & iCarB-Face & \textbf{0.3} & 1.3 & \textbf{0} & 0 \\
\hline
\multirow{3}{*}{iResNet100} & Multi-PIE & \textbf{0.1} & 0.5 & \textbf{0} & 0 \\
\cline{2-6}
 & SOTERIA & \textbf{0.2} & 0.7 & \textbf{0} & 0 \\
\cline{2-6}
 & iCarB-Face & \textbf{1.1} & 3.4 & \textbf{0} & 0.2 \\
\hline
\end{tabular}
\end{table} 

The aim of this experiment was to check whether key selection depends in some way on the structure of specific face embeddings (e.g., EdgeFace embeddings from the Multi-PIE dataset), or whether the keys can be selected independently of the face embeddings.  This experiment represents, therefore, a further check on the generalizability of the selected system-specific keys, this time not only across subjects (as in Section \ref{subsec:subj_spec}), but also across face recognition models (feature extractors) and datasets, i.e., checking whether keys selected from random vectors can be applied towards generating irreversible PolyProtected templates regardless of the feature extractor or dataset used to produce the face embeddings that are being protected using these keys.  From Fig. \ref{fig:expt_2} and Table \ref{tab:expt_2} we see that we can, indeed, select ``good'' system-specific keys (i.e., $C$ and $E$ parameters that produce \textit{irreversible} PolyProtected templates) by applying the key selection algorithm from \cite{khm26} to random vectors.  In fact, we see an almost perfect overlap between the two inversion score histograms (purple and blue) in Fig. \ref{fig:expt_2}, which indicates that the irreversibility of PolyProtected templates created using system-specific keys that come from random vectors is approximately the same as the irreversibility provided by system-specific keys selected from subject-specific keys.  The ISR values in Table \ref{tab:expt_2} confirm this observation, especially at the stricter threshold (0.01\% FMR).  At the looser threshold (0.1\% FMR), the ISR of PolyProtected templates whose system-specific keys come from random vectors (\textit{RV}) is consistently lower than the ISR of templates whose system-specific keys come from subject-specific keys (\textit{SK}).  This indicates that basing the key selection algorithm from \cite{khm26} on random vectors would actually result in the generation of slightly \textit{more irreversible} protected templates, compared to when the algorithm is first used to select subject-specific keys and system-specific keys are then chosen from this set; however, since the ISR in both scenarios (i.e., \textit{RV} and \textit{SK}) is quite low, in practice there may not be a significant difference in the degree of irreversibility provided by the two approaches, especially when the ISR is calculated at a strict threshold.  Having said that, the generation of system-specific keys from random vectors is the more practical approach.

Overall, the intriguing \textbf{new finding} from Fig. \ref{fig:expt_2} and Table \ref{tab:expt_2} is that the key selection algorithm from \cite{khm26} does not depend on the structure of the face embeddings that the selected keys are used to protect.  This means that it is possible to select system-specific keys that produce \textit{irreversible} PolyProtected templates for all subjects enrolled in the protected face recognition system, even without access to those embeddings.  We may thus conclude that, using this approach, it is possible to safely employ PolyProtect in the system-specific key scenario.  This is a useful finding, as it suggests the possibility of performing de-duplication of the enrolled face identities while the underlying face data remains protected.

\subsection{System-specific key selection \#3: From a C/E distribution}
\label{subsec:c_e}

Section \ref{subsec:rand_vec} proved that it is possible to choose ``good'' system-specific keys (i.e., $C$ and $E$ parameters that result in \textit{irreversible} PolyProtected templates) by applying the key selection algorithm from \cite{khm26} to random vectors.  This built upon the findings from Section \ref{subsec:subj_spec}, confirming the \textit{generalizability} of the key selection algorithm to \textit{system}-specific keys: we now know for sure that it is possible to select a single set of ($C$, $E$) parameters that would produce irreversible PolyProtected templates from the face embeddings of all subjects enrolled in a face recognition system.  This allows us to expand the operational capabilities of PolyProtect from the original \textit{subject}-specific key scenario (on which the key selection algorithm in \cite{khm26} was based) to the \textit{system}-specific key scenario, which is a promising step towards our goal of performing identity de-duplication in the protected domain.  

\begin{figure}[!h]
\centering
\includegraphics[width=0.9\columnwidth]{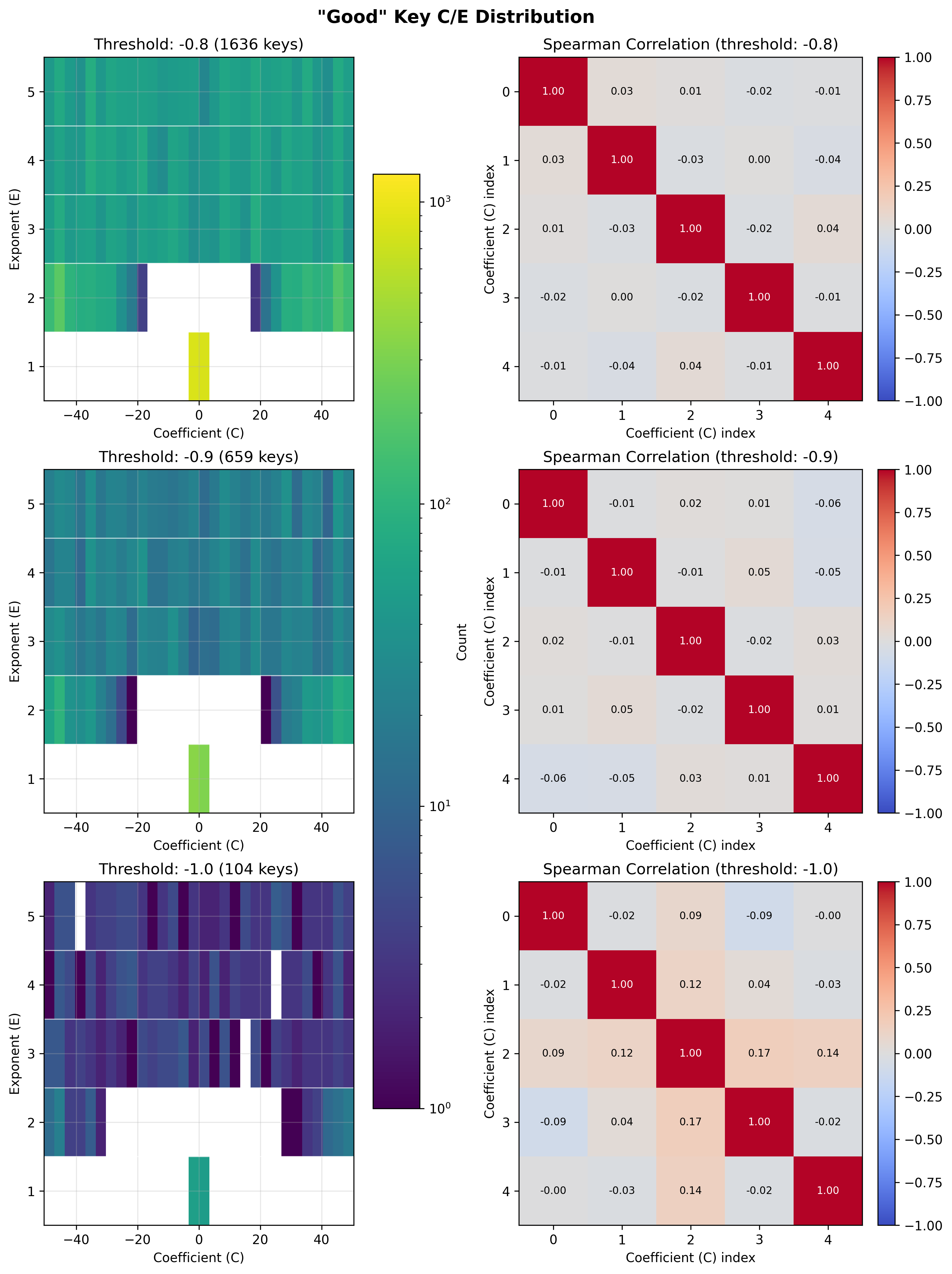}%
\caption{\textbf{Left column:} Distribution of ($C_i$, $E_i$) pairs from ``good'' keys selected by the key selection algorithm \cite{khm26}, when it was applied to random vectors and when different thresholds (negative cosine distance) were used to define irreversibility.  \textbf{Right column:} Spearman's rank correlation coefficient matrix, showing weak monotonic correlations across the different ($C_i$, $E_i$) pairs.}
\label{fig:c_e_distr}
\end{figure}

As a final test of the generalizability of the key selection algorithm from \cite{khm26} to the \textit{system}-specific key scenario, we performed a third experiment, which forms the subject of this section.  In particular, we wished to find out whether it would be possible to establish a representative \textit{distribution} of ``good'' ($C_i$, $E_i$) pairs, such that any future system-specific keys (i.e., sets of these pairs) could be selected directly from this distribution instead of from pre-generated 5-pair sequences (i.e., specific $C_1, C_2, ..., C_5$ and $E_1, E_2, ..., E_5$ sequences).  To conduct this experiment, we collected the entire set of ``good'' keys found within our 48-hour period for the random vectors from Section \ref{subsec:rand_vec}, which consisted of sequences of 5 corresponding $C$ and $E$ values.  We then used this set of keys to establish a distribution of ``good'' ($C_i$, $E_i$) pairs.  Fig. \ref{fig:c_e_distr} illustrates the resulting distribution, when different thresholds are used to define irreversibility in the key selection algorithm.  This distribution is accompanied by a Spearman's rank correlation coefficient matrix, which serves to unearth any monotonic relationships between different ($C_i$, $E_i$) pairs, thereby indicating whether certain pairs should (or should not) appear in the same ($C$, $E$) sequence in the selected keys.  

There are several interesting observations and \textbf{new findings} from Fig. \ref{fig:c_e_distr}.  Starting with the left column, we firstly see that the number of ``good'' keys (i.e., sequences of 5 ($C_i$, $E_i$) pairs found by the key selection algorithm and used to construct the illustrated $C$/$E$ distribution) decreases at stricter irreversibility thresholds: at a threshold of -0.8 the key selection algorithm found 1,636 suitable keys, while for a threshold of -1.0 this number dropped to 104.  This makes sense, because the lower (more negative) the threshold, the stricter the key selection algorithm becomes, making it more difficult to find sets of ($C$, $E$) parameters that would produce irreversible protected templates.  Secondly, we observe that, across all thresholds, the key selection algorithm is quite picky when choosing $C$ values for $E$ = 1 and $E$ = 2.  We note specifically that, for $E$ = 1, the corresponding $C$ is always chosen to be very small.  This makes sense, since an exponent of 1 corresponds to the linear term in the resulting PolyProtect polynomial, meaning that the key selection algorithm would try to reduce its effect to ensure that the protected template is irreversible (non-invertible).  As for $E$ = 2, we see that the $C$ value is always chosen to be large: the stricter the threshold (e.g., -1.0), the larger the $C$ value.  This also makes sense, because an exponent of 2 corresponds to the quadratic term in the PolyProtect polynomial, so the key selection algorithm would try to make it large in order to contribute a higher level of non-invertibility to the resulting protected template.  For the other $E$ values (3, 4, 5), we do not observe a specific pattern in the values of the chosen $C$ parameters.  This may be due to the fact that controlling for $E$ of 1 and 2 is already enough to ensure that the resulting PolyProtected template is irreversible.\footnote{We could try to make the key selection algorithm ``smarter'', so that it considers more carefully the contributions of \textit{all} polynomial terms, but for now we have sufficient evidence that it is good enough to generate \textit{irreversible} PolyProtected templates with both subject- and system-specific keys.}  Considering the right column of Fig. \ref{fig:c_e_distr}, we observe that the correlation coefficients are all small, most of them close to 0.  This tells us that there is no monotonic relationship (linear or non-linear) between the different ($C_i$, $E_i$) pairs in this study, meaning that each pair can be considered effectively independent of the other pairs.  This finding was important for designing the next step of this experiment, which involved constructing ($C$, $E$) sets from individual ($C_i$, $E_i$) pairs. 

\begin{figure*}[!h]
\centering
\subfloat{\includegraphics[width=0.3\textwidth]{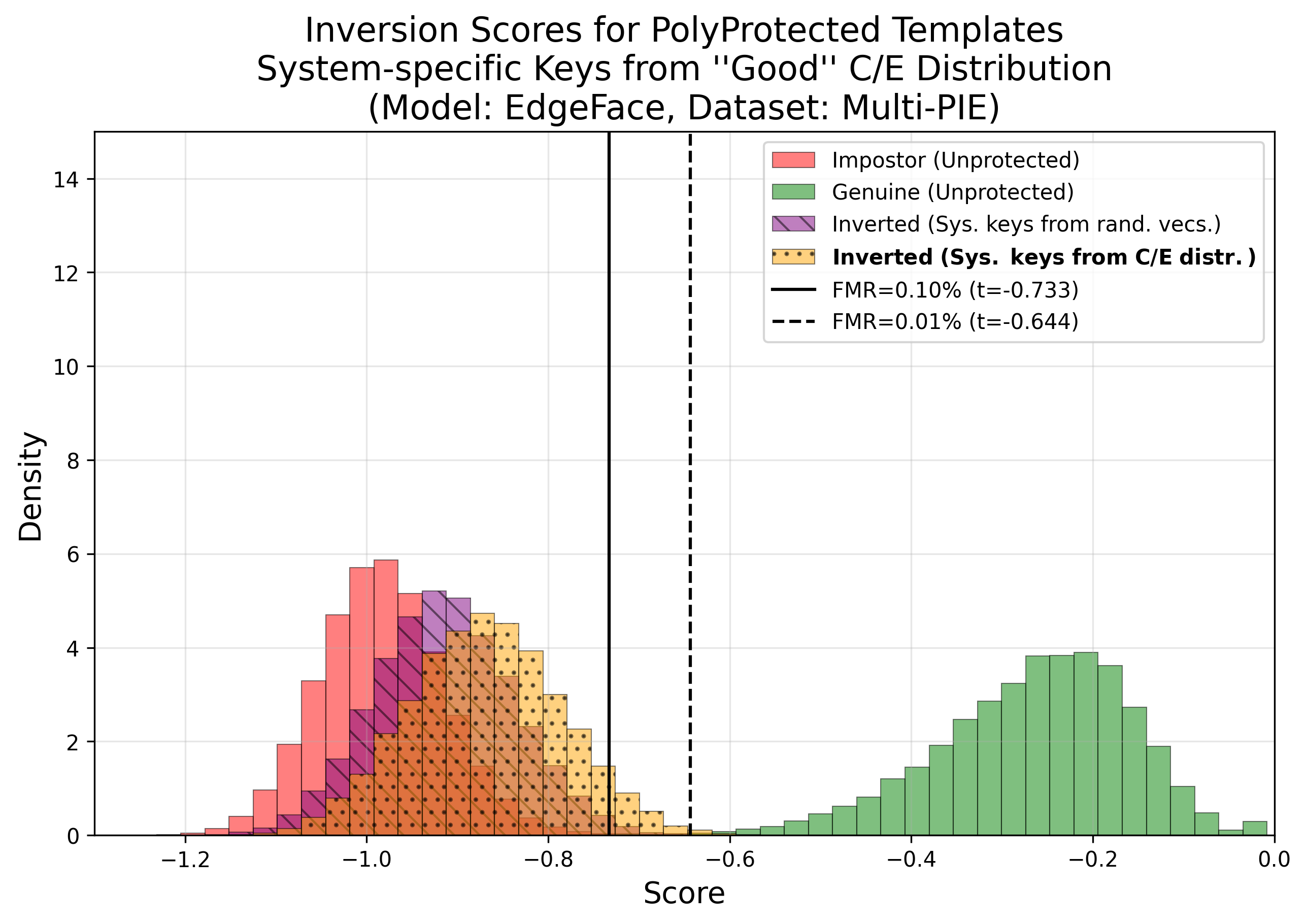}}
\hfil 
\subfloat{\includegraphics[width=0.3\textwidth]{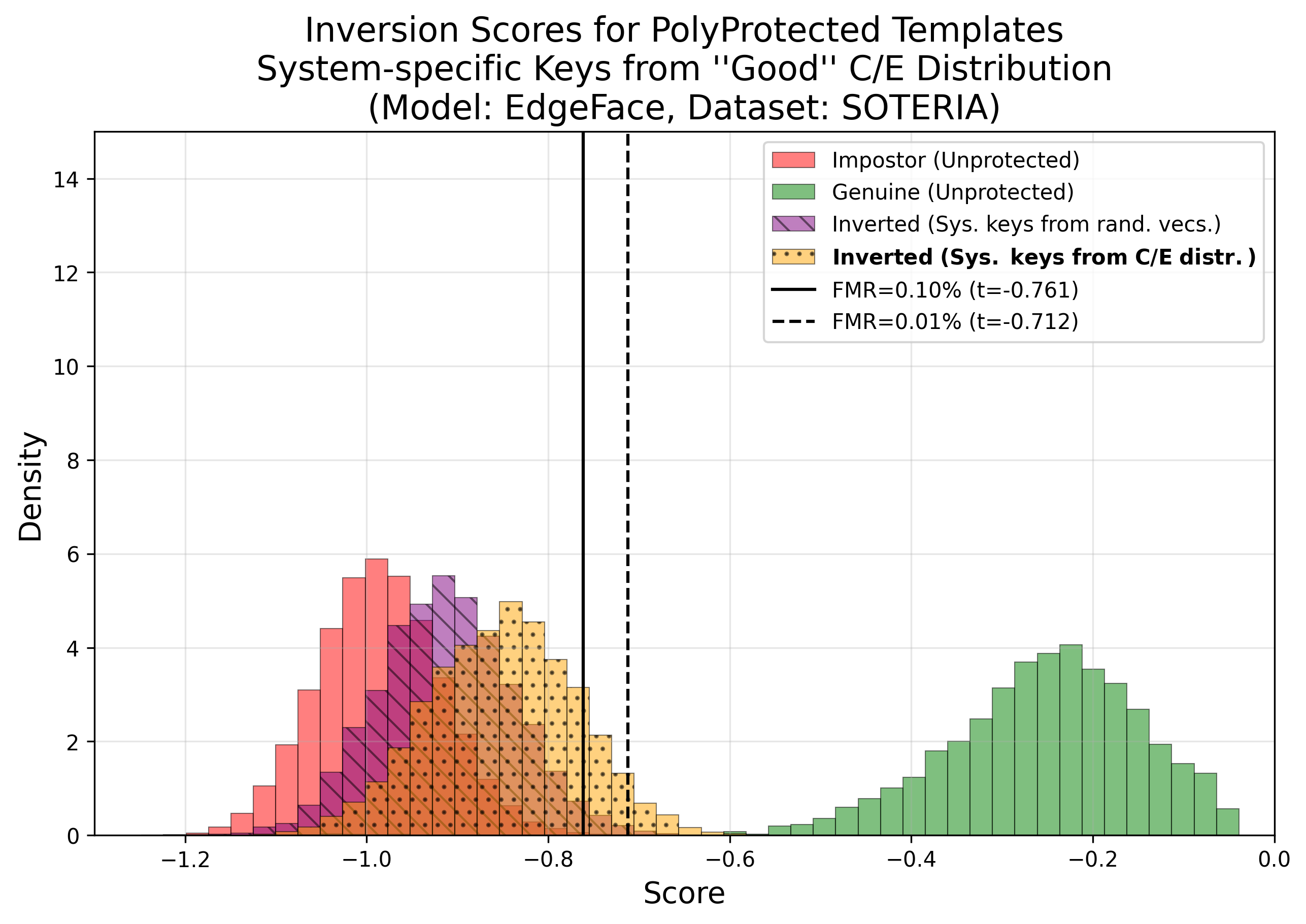}}
\hfil
\subfloat{\includegraphics[width=0.3\textwidth]{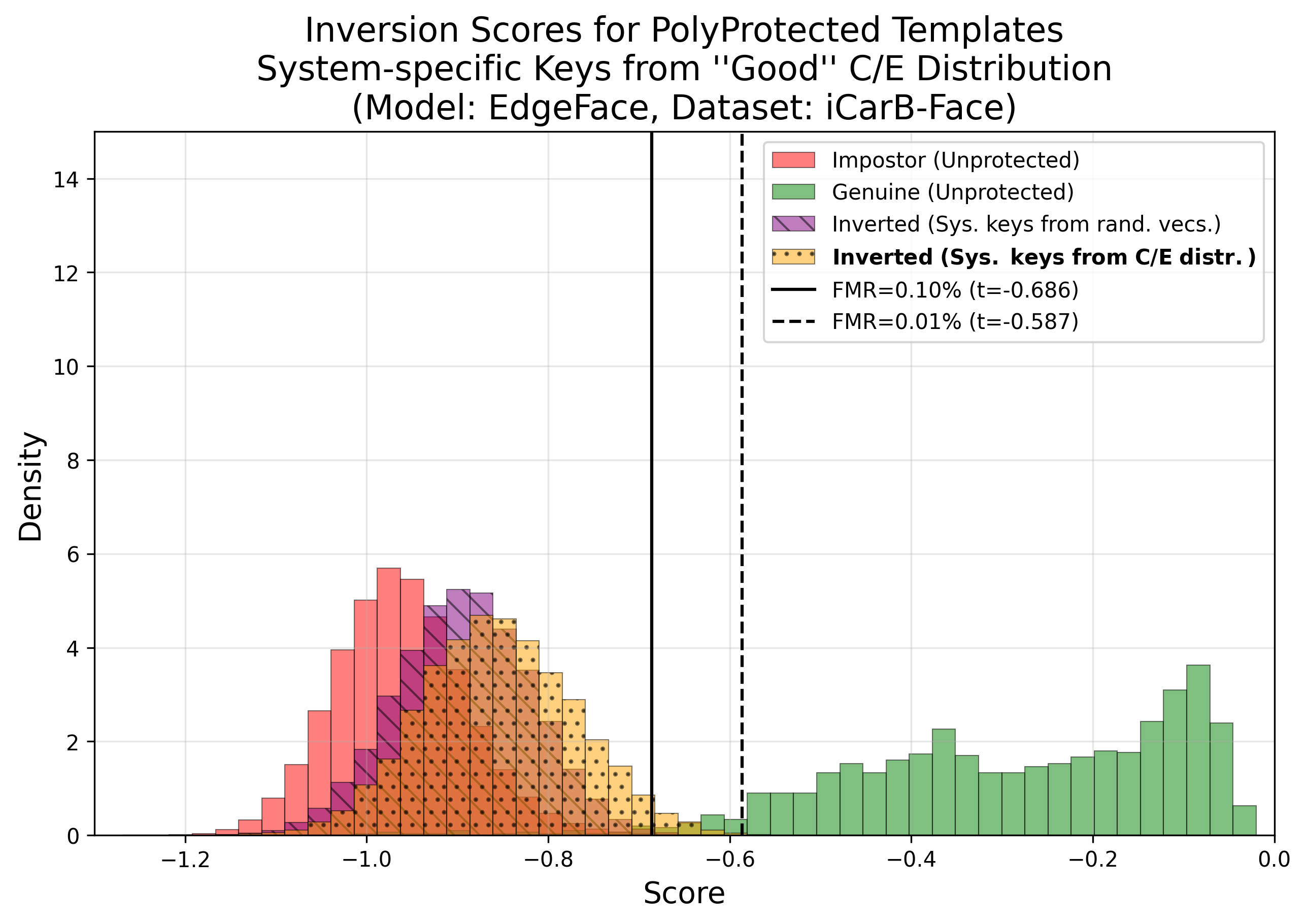}}
\vfil
\subfloat{\includegraphics[width=0.3\textwidth]{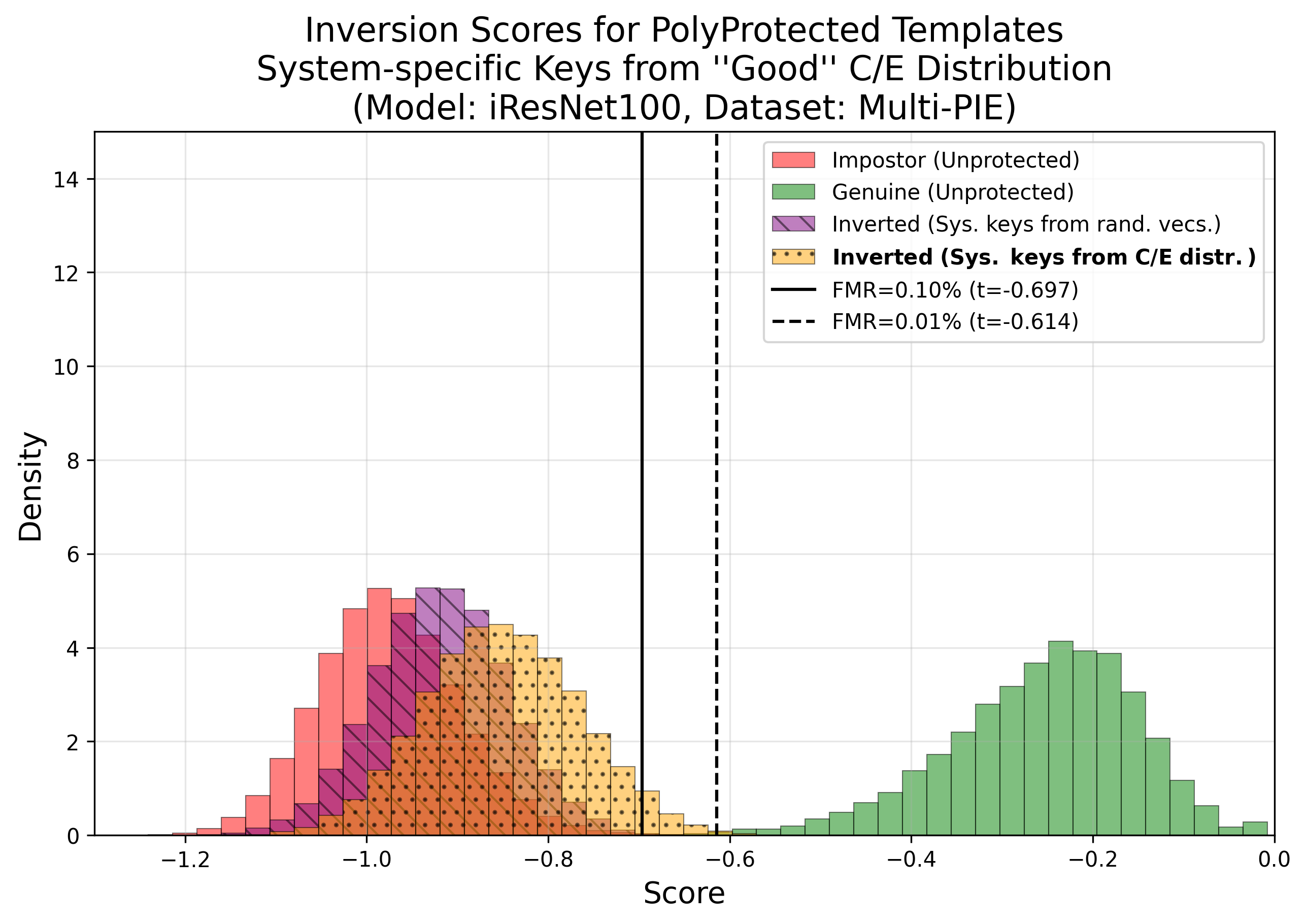}}
\hfil
\subfloat{\includegraphics[width=0.3\textwidth]{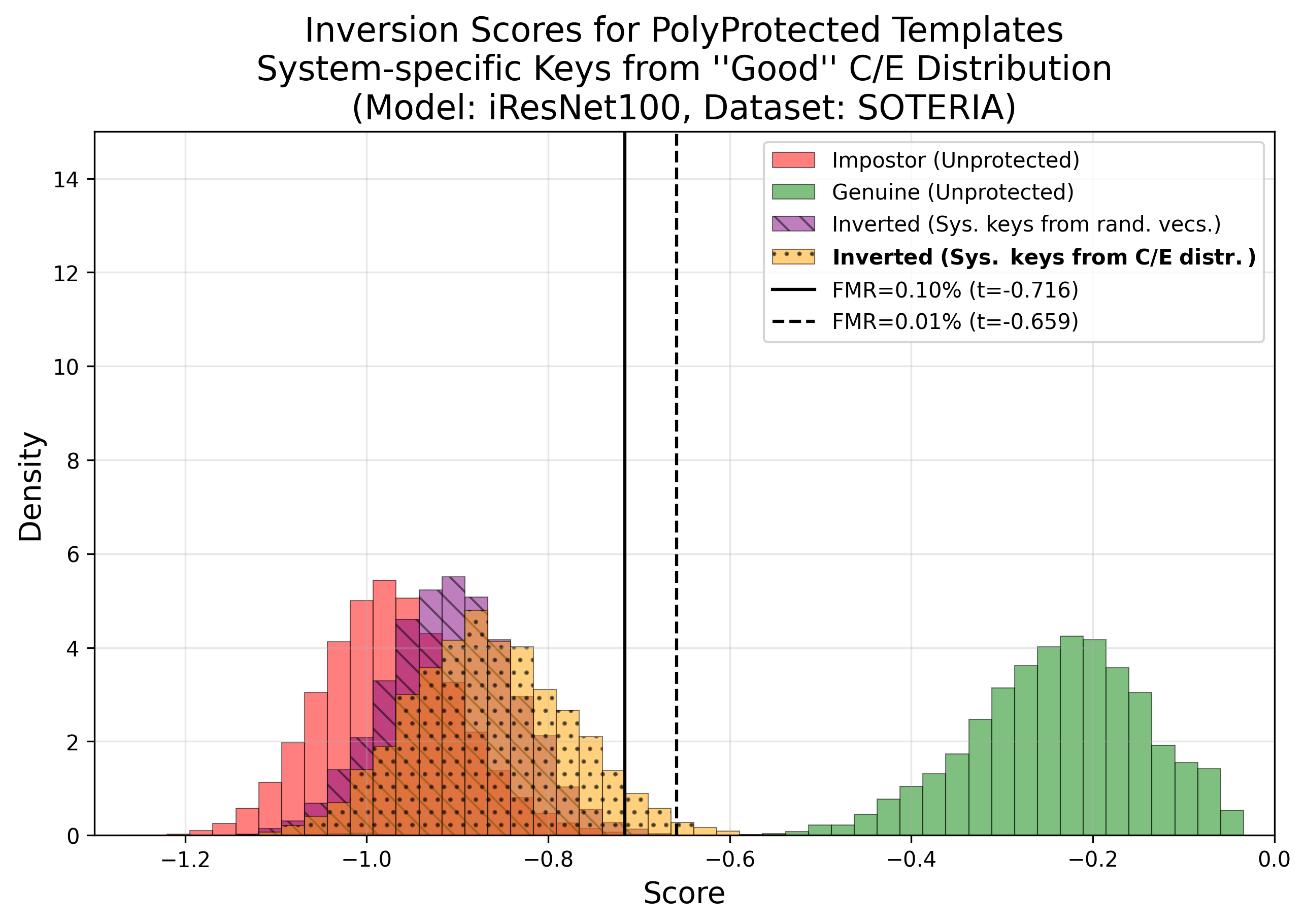}}
\hfil
\subfloat{\includegraphics[width=0.3\textwidth]{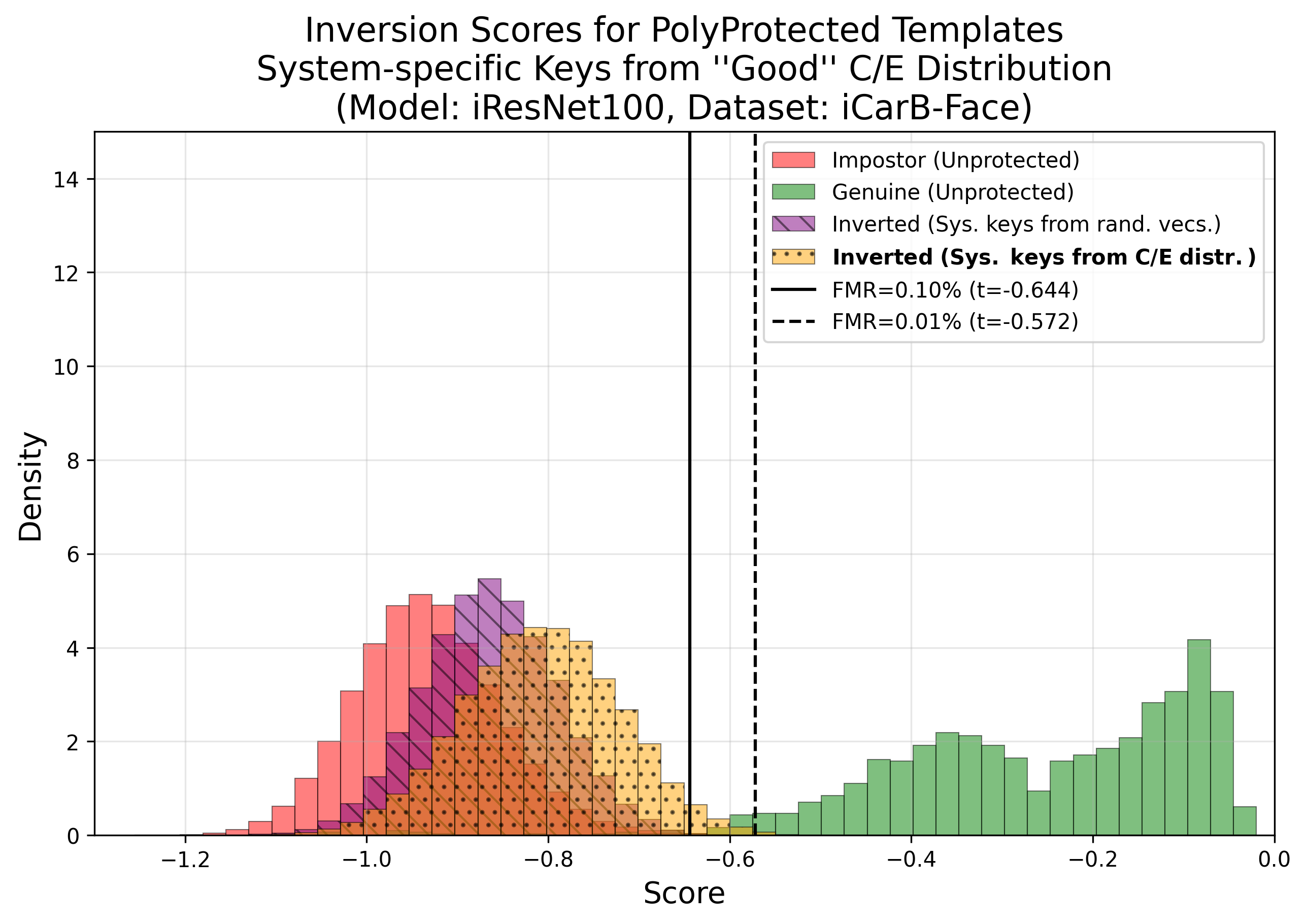}}
\caption{Inversion scores for PolyProtected templates generated from EdgeFace and iResNet100 face embeddings when the keys ($C$, $E$) were system-specific and based on applying the key selection algorithm from \cite{khm26} to random vectors.  The sequences of 5 corresponding $C$ and $E$ parameter pairs were: constructed by choosing samples from a ``good'' ($C_i$, $E_i$) pair distribution (orange histogram) versus chosen directly from a list of pre-determined sequences (purple histogram, from Fig. \ref{fig:expt_2}).  Our main focus is the \textbf{orange} histogram.  The inversion was performed using the numerical solver from \cite{khm26}.  The two vertical black lines represent different thresholds at which the inversion success rate (ISR) was computed for Table \ref{tab:expt_3}.}
\label{fig:expt_3}
\end{figure*}

Once we established our ``good'' ($C_i$, $E_i$) pair distribution (Fig. \ref{fig:c_e_distr}), we constructed 10 sets of 5 randomly chosen pairs from this distribution, i.e., ($C_1$, $E_1$), ($C_2$, $E_2$), ..., ($C_5$, $E_5$).  Since Fig. \ref{fig:c_e_distr} indicated that there is no correlation between different pairs of $C$ and $E$ parameters (within our set of pre-established ``good'' keys), each pair was (randomly) selected independently of the other pairs.  The resulting 10 sequences (of 5 pairs each) represented our system-specific keys in this experiment, and we selected a different set of 10 keys for each of our two sets of face embeddings (EdgeFace and iResNet100) from each of our three datasets.  Next, each of these 10 keys was applied, one at a time, to the corresponding face embeddings to generate the PolyProtected templates.  Finally, as in the experiments presented in Sections \ref{subsec:subj_spec} and \ref{subsec:rand_vec}, the cosine-solver-based inversion attack from \cite{khm26} was then launched on the protected templates to test their irreversibility.  Fig. \ref{fig:expt_3} compares the resulting inversion score histograms (based on the strictest considered irreversibility threshold of -1.0) to the histograms corresponding to the scenario where the system-specific keys are selected from the list of ``good'', ready-made sequences ($C_1, C_2, ..., C_5$ and $E_1, E_2, ..., E_5$) established on random vectors in Section \ref{subsec:rand_vec}.  Table \ref{tab:expt_3} compares the ISRs.
 
\begin{table}[!h]
\renewcommand{\arraystretch}{1.2}
\caption{Inversion success rate (ISR) for Fig. \ref{fig:expt_3} at the two FMR thresholds, for EdgeFace and iResNet100 face embeddings protected via PolyProtect when the keys (based on applying the key selection algorithm \cite{khm26} to random vectors) were system-specific and: \textbf{selected from a ``good'' ($C_i$, $E_i$) pair distribution (\textit{Dis.})} VS chosen from a list of pre-selected ($C$, $E$) sequences (\textit{Seq.} -- corresp. to \textit{RV} in Table \ref{tab:expt_2}). \label{tab:expt_3}} 
\centering
\begin{tabular}{|c|c|c|c|c|c|}
\hline
\multirow{3}{*}{Model} & \multirow{3}{*}{Dataset} & \multicolumn{4}{c|}{ISR (\%)} \\
\cline{3-6}
 & & \multicolumn{2}{c|}{@ 0.1\% FMR} & \multicolumn{2}{c|}{@ 0.01\% FMR} \\
\cline{3-6}
 & & \textbf{\textit{Dis.}} & \textit{Seq.} & \textbf{\textit{Dis.}} & \textit{Seq.} \\
\hline
\multirow{3}{*}{EdgeFace} & Multi-PIE & \textbf{2.1} & 0.3 & \textbf{0.1} & 0 \\
\cline{2-6}
 & SOTERIA & \textbf{2.6} & 0.2 & \textbf{0.2} & 0 \\
\cline{2-6}
 & iCarB-Face & \textbf{3.3} & 0.3 & \textbf{0.2} & 0 \\
\hline
\multirow{3}{*}{iResNet100} & Multi-PIE & \textbf{3.7} & 0.1 & \textbf{0.2} & 0 \\
\cline{2-6}
 & SOTERIA & \textbf{3.4} & 0.2 & \textbf{0.3} & 0 \\
\cline{2-6}
 & iCarB-Face & \textbf{10.1} & 1.1 & \textbf{1.1} & 0 \\
\hline
\end{tabular}
\end{table} 

The \textbf{novel finding} from Fig. \ref{fig:expt_3} and Table \ref{tab:expt_3} is that selecting system-specific keys in terms of individual $C$ and $E$ pairs from a ``good'' $C$/$E$ pair distribution is almost as effective as selecting system-specific keys from full ($C$, $E$) sequences generated by applying the key selection algorithm \cite{khm26} to random vectors (Section \ref{subsec:rand_vec}).  More specifically, in Fig. \ref{fig:expt_3} we see that the orange histogram overlaps significantly with the purple histogram, but is shifted slightly to the right.  This indicates that face embeddings protected using system-specific keys selected in terms of individual ($C_i$, $E_i$) pairs from a $C$/$E$ distribution may occassionally be easier to invert than protected templates generated using system-specific keys selected from pre-generated, full ($C$, $E$) sequences.  This is confirmed in Table \ref{tab:expt_3}, where the ISR of PolyProtected templates generated using the distribution-based key selection method (\textit{Dis.}) is generally higher than the ISR from using the sequence-based key selection method (\textit{Seq.}), with the difference being particularly noticeable at the looser threshold (0.1\% FMR).  This suggests that pre-generating system-specific keys on random vectors (i.e., generating the full $C_1, C_2, ..., C_5$ and $E_1, E_2, ..., E_5$ sequences) may be a more effective way of ensuring that the resulting PolyProtected templates are \textit{irreversible}.  This makes sense, since pre-generated ($C$, $E$) sequences have already been tested (on random vectors) to ensure that they produce irreversible protected templates, while sequences created using randomly selected pairs of coefficients and exponents from the $C$/$E$ distribution have not.  That said, Table \ref{tab:expt_3} shows that the difference in the ISR values between the \textit{Dis.} and \textit{Seq.} key selection approaches at the stricter ISR threshold (0.01\% FMR) is quite small.  This suggests that selecting system-specific PolyProtect keys from pre-established $C$/$E$ distributions may be sufficiently effective for ensuring that the face embeddings protected using these keys are irreversible, especially if the ISR is computed at stricter thresholds.  Furthermore, we observed that using a smaller ``overlap'' for the PolyProtect transform also tends to reduce the ISR difference between the two key selection approaches.\footnote{Recall that in this paper we are using overlap = 3.  Results for other overlaps can be generated using our open-source code.}  Having said that, if pre-generation of full ($C$, $E$) sequences, by applying the key selection algorithm \cite{khm26} to random vectors, is possible for a given application in practice, this seems to be an overall safer way to select system-specific keys for the PolyProtect BTP method (in terms of ensuring a lower ISR, regardless of the ``overlap'' used in the transform).

\subsection{How do system-specific keys affect recognition accuracy?}
\label{subsec:accuracy}

Section \ref{subsec:motivation} motivated our focus on \textit{system}-specific PolyProtect keys, from the point of view of enabling face identity de-duplication in the protected template domain.  Sections \ref{subsec:subj_spec} -- \ref{subsec:c_e} then presented three experiments to prove the generalizability of the key selection algorithm \cite{khm26} from the subject-specific to the system-specific key scenario.  These experiments showed that it is possible to select system-specific keys that can generate \textit{irreversible} PolyProtected templates across all subjects enrolled in a face recognition system.  The question we have now is the following: How do the selected system-specific keys affect the \textit{accuracy} of the face recognition system?  We know from \cite{khm26} that the proposed key selection algorithm, when used for the selection of \textit{subject}-specific keys, results in some loss of accuracy in the protected domain, compared to when the PolyProtect keys are selected randomly.  In this section, we investigate whether the same applies to the selection of \textit{system}-specific keys.

\begin{figure}[!h]
\centering
\subfloat{\includegraphics[width=0.33\columnwidth]{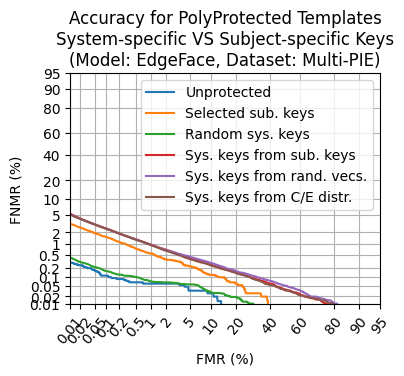}}
\hfil 
\subfloat{\includegraphics[width=0.33\columnwidth]{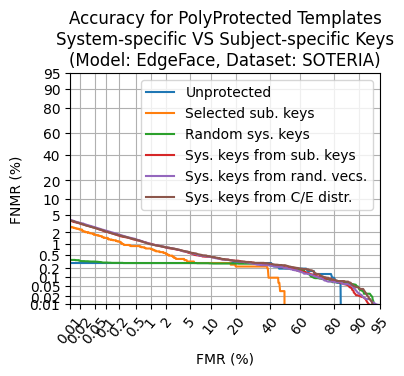}}
\hfil
\subfloat{\includegraphics[width=0.33\columnwidth]{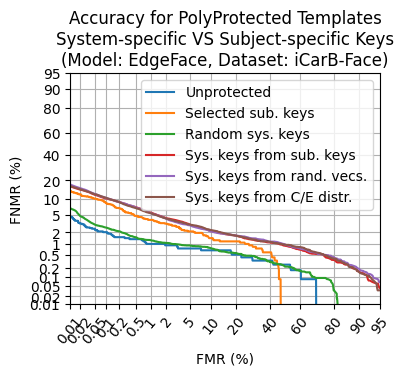}}
\vfil
\subfloat{\includegraphics[width=0.33\columnwidth]{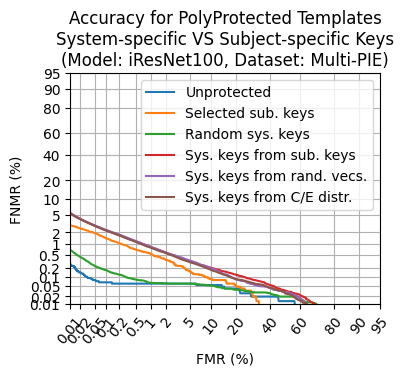}}
\hfil 
\subfloat{\includegraphics[width=0.33\columnwidth]{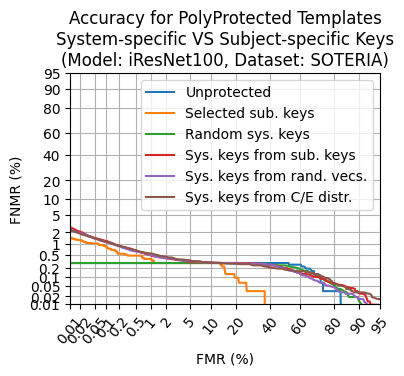}}
\hfil
\subfloat{\includegraphics[width=0.33\columnwidth]{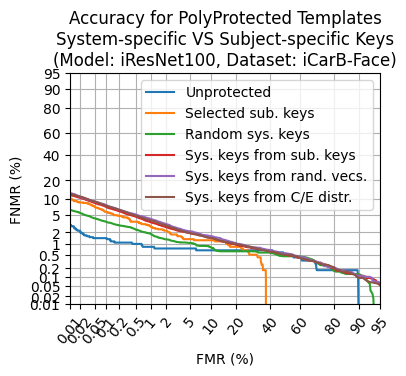}}
\caption{Recognition accuracy for PolyProtected templates when the keys ($C$, $E$) were generated randomly versus using the key selection algorithm from \cite{khm26}.  In the latter scenario, the keys were either subject-specific, or system-specific and selected using the three methods from Sections \ref{subsec:subj_spec} -- \ref{subsec:c_e}: from subject-specific keys, from random vectors, or from a $C$/$E$ distribution.  Unprotected template accuracy is shown as a baseline.}
\label{fig:accuracy}
\end{figure}

Fig. \ref{fig:accuracy} compares the recognition accuracy in the protected domain when the PolyProtect keys are chosen using the key selection algorithm from \cite{khm26} versus when they are generated randomly.  In the former scenario, we compare \textit{subject}-specific keys to \textit{system}-specific keys selected using the three methods explored in our work (see Sections \ref{subsec:subj_spec}, \ref{subsec:rand_vec}, and \ref{subsec:c_e}).

The first observation from Fig \ref{fig:accuracy} is that, for system-specific key selection, the recognition accuracy in the protected domain remains approximately the same regardless of which of the three key selection methods investigated in this work (explored in Sections \ref{subsec:subj_spec}, \ref{subsec:rand_vec}, and \ref{subsec:c_e}) is used.  The second observation is that selected \textit{system}-specific keys result in slightly worse accuracy in the protected domain, compared to selected \textit{subject}-specific keys.  This may be attributed to the fact that, in the subject-specific key scenario, the unique $C$ and $E$ parameters used to define the PolyProtect transform for each subject (identity) effectively act as a second authentication factor, making it easier to distinguish between the protected templates of different subjects.  In the system-specific key scenario, on the other hand, all subjects' face embeddings are transformed in exactly the same way (using the same $C$ and $E$ parameters), so there is no extra, external factor to contribute to the separation of identities in the protected domain.  Nevertheless, from Fig. \ref{fig:accuracy} it is clear that the accuracy drop from subject-specific to system-specific keys is not large, so this may be considered a reasonable trade-off for the additional de-duplication capability provided by system-specific keys.  The third observation from Fig. \ref{fig:accuracy} is that using the key selection algorithm \cite{khm26} to choose either system-specific or subject-specific keys results in worse recognition accuracy in the protected domain compared to when these keys are generated randomly.  This was already noted for the subject-specific key scenario in \cite{khm26}, and Fig. \ref{fig:accuracy} confirms that the same observation holds for the system-specific key scenario.  This seems a reasonable compromise considering the significantly more important improvement in irreversibility (i.e., decrease in ISR from \textit{R-Sy.} to \textit{Sy.} in Table \ref{tab:expt_1}).  Boosting the accuracy should nevertheless be considered as the next step towards ensuring effective de-duplication in the protected domain.

\section{Conclusions and Future Work}
\label{sec:conclusion}

This paper investigated the possibility of generating \textit{irreversible protected templates} from face embeddings by employing the PolyProtect BTP method with \textit{system}-specific keys, instead of the traditional \textit{subject}-specific keys.  This work was motivated by the ultimate goal of performing face identity de-duplication in the protected template domain, which was shown to be impossible when subject-specific keys are used to create the protected templates (due to PolyProtect's unlinkability property), while system-specific keys were found to be suitable for this purpose.  The question we were interested in was the following: Is it possible to select system-specific keys (i.e., a single set of $C$ and $E$ parameters defining the polynomial transform for all subjects enrolled in a face recognition system) such that the resulting PolyProtected templates would be \textit{irreversible}?  We presented three experiments proving that this question can be answered in the affirmative, when system-specific keys are chosen in the following ways: (i) from subject-specific keys selected using the key selection algorithm proposed in \cite{khm26}, (ii) by applying the key selection algorithm to random vectors, and (iii) by approximating a ``good'' $C$/$E$ pair distribution, from which system-specific keys can be constructed in terms of sequences of suitable $C$/$E$ pairs.  Overall, therefore, we established at least three ways of successfully selecting system-specific keys that, when applied to face embeddings, lead to the generation of \textit{irreversible} PolyProtected templates.  Although the recognition accuracy provided by these templates was found to be lower than when the keys are generated randomly, the corresponding improvement in irreversibility was deemed much more significant, indicating an acceptable trade-off.  The implication of these findings is that it is possible to operate PolyProtect in the system-specific key scenario while guaranteeing protection of the underlying face data, which provides a legitimate avenue for identity de-duplication in the protected domain.  Designing a robust de-duplication method is, therefore, the natural next step.  To effectively detect duplicates, this method will need to ensure sufficient recognition accuracy among the protected templates.  We plan to investigate three approaches: (i) controlling the desired accuracy/irreversibility compromise, by tuning the irreversibility threshold in the key selection algorithm or enhancing the algorithm to choose system-specific keys that explicitly optimise this balance, (ii) incorporating more distinguishing information into the templates to facilitate the separation of protected identities by adopting multimodal biometrics (e.g., face + fingerprint), and (iii) considering hybrid BTP methods (e.g., PolyProtect + another technique) to try to alleviate the accuracy/irreversibility trade-off.

\section*{Acknowledgments}
This work was funded by the Innosuisse project ``PRiMEAiD: Privacy-pReserving bioMetric idEntification for humAnitarian aid Distribution'' (Number: 116.346 IP-ICT).

% References
\bibliographystyle{IEEEtran}
\bibliography{bibliography}

\vfill

\end{document}